\documentclass[preprint,aps,tightenlines,floatfix,showpacs]{revtex4}

\usepackage{graphics}
\usepackage{bm}
\usepackage{amsmath}
\usepackage{amssymb}

\begin{document}

\

\title{Relativistic interactions for the meson-two-nucleon system in the
clothed-particle unitary representation}
\author{V.Yu. Korda$^1$}
\email{
kvyu@kipt.kharkov.ua}
\author{L. Canton$^2$}
\email{
luciano.canton@pd.infn.it}
\author{A.V. Shebeko$^3$}
\email{
shebeko@kipt.kharkov.ua}



\affiliation
{$^1$Institute of Electrophysics and Radiation Technologies,
National Academy of Sciences of Ukraine, \\
28 Chernyshevsky St., P.O. BOX 8812, UA-61002 Kharkov, Ukraine }
\affiliation
{$^2$ Istituto Nazionale di Fisica Nucleare, Sez. di Padova, via F.
Marzolo, 8, I-35131 Padova, Italy}
\affiliation
{$^3$NSC Kharkov Institute of Physics and Technology, National
Academy of Sciences of Ukraine,
1 Akademicheskaya St. UA-61108 Kharkov, Ukraine }

\begin{abstract}
{\ The method of unitary clothing transformations put forward in
relativistic quantum field theory (QFT) by Greenberg and Schweber and
developed by Shirokov is applied to construct a new family of interactions
in the meson-two-nucleon system. Along with a brief exposition of its basic
elements we show a specific transition from the initial ``bare'' one-meson
and one-nucleon operators and states to their physical ``clothed''
counterparts. We emphasize that the clothing transformations in question do
not alter the original total Hamiltonian, but provides a conceptually more
transparent representation of the same Hamiltonian in terms of a new set of
operators for particles with physical properties and their relativistic
interactions. The Hermitian and energy-independent interaction operators for
the processes $\pi N\rightarrow \pi N$, $NN\rightarrow NN$ and $%
NN\leftrightarrow \pi NN$ are derived starting from the Yukawa-type
couplings between fermions (nucleons and antinucleons) and bosons ($\pi -$, $%
\eta -$, $\rho -$, $\omega -$ mesons, etc.). These types of interaction have
a distinctive off-energy-shell structure which is naturally generated by the
unitary transformation that removes from the Hamiltonian the (three-leg) $%
\pi $NN vertex coupling.}
\end{abstract}


\maketitle

\section{ Introduction}

To view the light hadronic systems as few-body systems involves several
problematic aspects, and many of these aspects still need to be
theoretically explored in detail. Not all cases are so fortunate that one
can accurately reduce the underlying field-theoretical problem to a
non-relativistic quantum-mechanical Hamiltonian formulation in a (two-body,
three-body, etc.) potential scheme. One fortunate situation occurs typically
in low-energy few-nucleon physics~\cite{Aren}, where the considered degrees
of freedom are the nucleons, and the non-relativistic interactions are
generated via meson-exchange processes, or via alternative, effective
field-theoretical derivations. But, even in this well-defined research area,
there have been attempts to brake the paradigm of the purely nucleonic
Hamiltonian description in the potential scheme~\cite{Canetal}, and to
relate the origin of well-known unsolved problems, such as the $A_y$ puzzle,
to dynamical effects of the pion~\cite{CanScha}.

As the energy increases, the light hadronic systems become much more
problematic to be described. The opening of the pionic threshold forces to
deal directly with the pion degrees of freedom. This complicates the
few-body equations since one has to deal simultaneously with phenomena of
production and absorption of particles~\cite{garzila-mizuta}. The derivation
of non-perturbative dynamical equations for meson-few-nucleon processes is
still waiting for a consistent theory that solves the general problem in a
feasible and calculable way.

Specific topics, such as the study of $NN\rightarrow NN\pi $ inelasticities,
both in free space and in more complex nuclear environments, have received
much attention recently. Many phenomenological works have attempted the
description of the experimental data, including polarization data, and
extensive reviews on the past and current status of research on this subject
are available\cite{garzila-mizuta,haiden,hanardt,Dubna1995,Cracow1996}. In
general, such models perform some nonrelativistic reduction of the
interaction Hamiltonian between mesons and nucleons, and construct the basic
production mechanisms using perturbative schemes. Typically, these terms
consist of the $\pi $NN one-body term (direct term plus recoil) and some
more complex two-nucleon processes such as pion rescattering, intermediate $%
\Delta $ excitations, and 2N processes involving heavy-meson exchange
currents. As yet, there is not a clear-cut understanding of which mechanisms
contribute the most and, especially in the isoscalar channel at threshold
(the channel of the pp$\rightarrow $pp$\pi ^o$ process) there has been quite
a debate on which are the basic mechanisms that provide the strength to the
threshold cross section. This strength can be generated either by
short-range heavy-meson exchanges~\cite{lee-riska}, or by pion rescatterings
in the $\sigma $ channel~\cite{hernandez-oset}. Presently, the analysis of
the $NN\rightarrow NN\pi $ inelasticities did not succeed to resolve this
ambiguity, and there is hope that the study of more complex reactions, such
as $pd\rightarrow \pi ^o\ He^3$, can help to solve this ambiguity.
Phenomenological studies for this reaction have been limited to calculations
involving only pion rescattering in the $\sigma $-channel~\cite
{Canton-Schadow}; these calculations achieved, very recently, a good
reproduction of data for both angular distributions for the excitation
function and for the analyzing power~\cite{Canton-Levchuk}. For this 3N
reaction, a thorough study involving production mechanisms originated also
by 2N heavy-meson exchanges is needed, but unfortunately these studies have
still to come.

Besides the problem of constructing the 2N heavy-meson exchange mechanisms
in a consistent manner (we will come back to this problem in the final part
of this paper), there remain, in addition, several difficulties that need to
be explored. One is related to the need to use relativistic equations for
the nucleon dynamics in the energy range of meson production, since this
energy range is higher than the typical low-energy nuclear regime where
nonrelativistic quantum mechanics works so well. Then, the first step is to
construct consistent relativistic (pseudo) potentials that fulfill the
Poincar\'e algebra.

Another difficulty is connected with the treatment of virtual effects
generated by the $\pi$NN vertex in the production process. To directly use
the one-body vertex and similar operators in the construction of
one-body-type production mechanisms may create serious difficulties because
in the $\pi NN$ vertex the three particles cannot stay simultaneously
on-mass-shell. However, the use of quantum-mechanical equations for the
nucleon dynamics forces the nucleons on their mass shell, and the pion,
being the physical observed particle in the production process, has also to
stay on its mass shell. Clearly, all this creates ambiguities in the
possible treatment of one-body operators, and hence there are problems of
internal inconsistencies in the construction and calculation of sets of
one-body, two-body, etc, operators. This represents a not negligible problem
in pion production at threshold: although the magnitude of the one-body term
is generally small, the interference effects of the one-body term with
respect to the more dominant two-body amplitudes have important consequences
in the cross sections.

A similar difficulty also occurs when the elementary fermion-meson vertex 
interaction is used for the construction of meson decays operators
in relativistic constituent quark models. It has been shown recently
that the use of the spectator-model ansatz (built upon the elementary
quark-meson vertex) in Relativistic Quantum Mechanics (RQM) is not fully
constrained in covariant calculations of strong (meson) decays of baryon
resonances. This aspect has been pointed out in Ref.~\cite{MelCanPleWag}
specifically for the construction of spectator-model operators in
the so-called Point-Form formulation of RQM. However, the presence of
ambiguities in the general construction of Current Operators 
has been observed in all forms of RQM, including front form and 
instant form~\cite{Lev95}.

The approach illustrated in this paper represents a field-theoretic way to
overcome the above mentioned difficulties. Among the varieties of
field-theoretical approaches we concentrate on the Unitary-Transformation
(UT) methods for Hamiltonian-based models. UT's do not change the
expressions for the observables, because the $S$-matrix remains the same
under such transformations. However, the unitary transformations allow to
change the representation of the Hamiltonian so that the interaction
operators refers to particles with physical (observable) properties. It is
important that the interactions extracted this way are Hermitian,
energy-independent and include the off-shell structures naturally. These
attractive features of the unitary transformation approaches allow to derive
pion-production interaction operators that are relativistic and involve only
physical particles (and hence, one-body operators are eliminated from the
representation and all the virtual processes or effects are embedded in the
interaction operators); we also extend the calculation to include 2N
heavy-meson exchange mechanisms for the production amplitude, given the
importance that these mechanisms have in discussing the phenomenology of 3N
forces and pion production.

There have been quite different implementations of UT techniques (see Ref.~%
\cite{SheShi2001} and refs. therein, for a review). In such approaches one
starts with the original field Hamiltonian $H$, to arrive at a new
representation of $H$ which is unitarily equivalent and where the particles
involved interact through interaction operators that are energy independent.
Such energy independence simplifies the solution of the eigenvalue problem
for the initial Hamiltonian. To avoid the confusion that already exists in
the literature, we do not employ here the inflated term ``effective'' to
qualify our interactions, since they have no direct connections (that we are
aware of) with the formalisms of effective Lagrangians.

Amongst the different UT formalisms, we mention first Okubo's idea \cite
{Okubo.Progr.Theor.Phys.12.603.1954} to block-diagonalize $H$ via a UT with
respect to the decomposition of the full Fock space $\mathcal{R}_F$ of
hadronic states into the nucleon (no-meson) subspace and its complement in $%
\mathcal{R}_F $. The Okubo approach was further developed by Gl\"{o}ckle and
M\"{u}ller \cite{GlockleMuller.Phys.Rev.C23.1183.1981} in their relativistic
theory of interacting particles. These authors were the first to show, al
least to our knowledge, how with the help of one single UT the non-commuting
Hermitian operators (Hamiltonian and Lorentz boosts) can be reduced to a
block form. For a simple model of ''spinless nucleons'' exchanging scalar
mesons, it was then possible to construct the new Hermitian and energy
independent contributions to the Poincar\'{e} group generators.

In Ref. \cite{KorchinShebekoPhysAtNucl1993} the same method was employed for
deriving the nucleon-nucleon and nucleon-antinucleon interactions starting
from a field Hamiltonian with the exchange of {$\pi -$, $\rho -$, $\omega -$
and $\sigma -$} mesons. Then, in the framework of the Hartree approximation,
these interactions have been introduced to describe the saturation
properties of nuclear matter. That approximation, we note in passing, gives
rise to an original recipe (see Eqs. (23)-(25) in \cite
{KorchinShebekoPhysAtNucl1993}) for calculating the nucleon mass shift in
the three dimensional approach.

Fuda and co-workers \cite{Fuda.Ann.Phys.231.1.1994}, \cite
{FudaZhang.Phys.Rev.C51.23.1995}, \cite{FudaZhang.Phys.Rev.C54.495.1996},
\cite{ElmessiriFuda.Phys.Rev.C60.044001.1999} have also used the
Okubo-Gl\"{o}ckle-M\"{u}ller approach to construct one-boson-exchange models
in light-front dynamics. They prefer to work with the mass-square operator $%
M^2$ rather than with $H$ because the former commutes with all of the
Poincar\'{e} generators. 
Such studies demonstrated that the two-particle interactions obtained in
second-order perturbation theory have certain similarities but do not
coincide exactly with the relevant Born terms obtained using the
Feynman-diagram techniques (see also Ref.~\cite{KorchinShebekoPhysAtNucl1993}%
). We shall address this question in more detail in Sec. 3.

Successful applications of the UT method have been obtained also in the
theory of nuclear forces and nuclear electromagnetic processes (see Refs.~%
\cite{SatoTamuraNiwaOhtsuboJ.Phys.G.17.303.1991}, \cite
{KobayashiSatoOhtsuboProgrTheorPhys1997} with reference to earlier studies
of the 50's, \textit{e.g.}, Ref.~\cite{Nishijima1956}). Such developments by
the Osaka group~\cite{KobayashiSatoOhtsuboProgrTheorPhys1997} illustrate
another useful application of the UT formalism: the elimination from $H$ of
certain virtual processes whose $S$-matrix elements, by definition, have to
vanish on the energy shell. Remarkably, the transformed Hamiltonian $%
H^{\prime }=U^{\dagger }HU$ contains only interactions for real processes.
Even more remarkably, it was possible to derive the one-pion-exchange (OPE)
and two-pion-exchange (TPE) nucleon-nucleon potentials, the pion-nucleon
potential (see also \cite{SatoLee.Phys.Rev.C54.2660.1996}), the $N\Delta
\rightarrow NN$ transition potential, the interactions for pion
absorption/production, and so on, from one and the same initial Hamiltonian $%
H$ without any additional \textit{ad hoc} assumption (of physical or
mathematical nature) in each case. The corresponding Hermitian
energy-independent operators are defined in the complete $\mathcal{R}_F$,
\textit{i.e.}, not only in a subspace $\subset \mathcal{R}_F$ as was the
case for Okubo approach.

In the present work, we construct the meson-baryon interaction operators
relying upon the notion of ''clothed'' particles. Such notion was introduced
many years ago by Greenberg and Schweber~\cite{GRESCH58} (see also Chapter
XII in the monograph~\cite{SchweberBook}) in their aim to include the
so-called cloud or persistent effects (the terminology refers back to Van
Hove \cite{vanHove.Physica.21.901.1955}, \cite{vanHove.Physica.22.343.1956}%
). In this connection, we apply the method of unitary clothing
transformations elaborated by \cite{FaddeevDokl.AkadNauk.USSR.1963.152.573}
and \cite{ShirokovVisinescuRevRoumPhys1974} and developed recently in Refs.
\cite{SheShi97}, \cite{SheShi98}, \cite{SheShi2000}, \cite{SheShi2001}. In
this way, a large amount of virtual processes induced by the meson
absorption/emission, the $N\bar{N}$ -pair annihilation/production and other
cloud effects can be accumulated in the creation (destruction) operators for
the clothed (physical) mesons and nucleons. Such a bootstrap reflects the
most significant distinction between the concepts of clothed and bare
particles.

The approach used in this article differs from the aforementioned ones at
least in two aspects. First, the clothing procedure does not aim, \textit{a
priori}, to define a UT that blockdiagonalizes $H$, as is done in the Okubo
approach. Besides, it might be not always possible to implement such a task
in the infinite-dimensional Hilbert space. Instead, the aim of the multistep
clothing procedure is to express the original Hamiltonian $H$ in terms of
the new clothed-particle operators in a form which is different than that
given for the initial bare-particle ones. Such a transition from the
bare-particle representation (BPR) to the clothed-particle representation
(CPR) introduces a new sparse structure of the original Hamiltonian $H$.
Second, in the framework of the CPR the mass and vertex renormalization
problem~\cite{SheShi2001},~\cite{KordaShebekoPhys.Rev.2004} is considered in
a natural way, in parallel with the construction of the interactions, while
the renormalization problem in Ref.~\cite
{KobayashiSatoOhtsuboProgrTheorPhys1997} has been disregarded. Other
distinctions are discussed in Sects. 2 and 3.

In this context, we would like to draw attention to the recent work by
Stefanovich on the problem of divergences in QFT. As was shown in Ref.~\cite
{StefanovichAnnPhys2001}, one can introduce a similarity transformation $%
U^{\dagger }HU$ to cancel infinite counterterms directly in $H$ and
therefore also in the $S$-matrix operator. As an application, the method has
been illustrated for the case of an Hamiltonian system in quantum
electrodynamics. Although Ref.~\cite{StefanovichAnnPhys2001} does not use
the clothing ideas, there are many points of contact between that approach
and the CPR method and, in our opinion, further developments in this area
could have promising perspectives.

The central goal of this paper is the derivation of interaction operators in
the CPR for the processes $\pi N\rightarrow \pi N$, $NN\rightarrow NN$ and $%
NN\leftrightarrow \pi NN$. For the field-theoretical treatment of mesons and
nucleons we assume the instant form after Dirac of relativistic dynamics, in
which the generators of space translations and rotations are the same as in
the free theory. We remind that in this case the three boost operators must
involve the interaction parts to meet, together with the total Hamiltonian $%
H $, the well-known commutation relations of the Poincar\'{e}-Lie algebra.
At the beginning we prefer to consider $H$ as a function of creation and
destruction operators in the Fock space rather than the more customary form
that follows from a given field Lagrangian. Our option follows a general
statement that $H$ may be expressed as a sum of products of creation and
destruction operators (see Chapter III of Ref.~\cite{WeinbergBook1995}).
This Hamilton formulation of RQFT simplifies the introduction of the CPR
that we consider in this work.

The outline of this paper is as follows. The aspects of the method of
unitary clothing transformations, which are necessary for constructing a new
family of interactions operators in the meson-two-nucleon system are
illustrated in Sec. 2. We introduce an auxiliary UT's that convert the
primary bare bosons and fermions into some intermediate-level particles with
physical masses. The approach develops along the chain: bare particles with
bare masses $\to $ bare particles with physical masses $\to $ physical
(observable) particles. This procedure is useful for drawing some parallels
between the clothing approach in QFT and the method of canonical
transformations (in particular, the Bogoliubov ones) in the theory of
superfluidity and superconductivity. Analytical expressions for the
relativistic interactions (quasipotentials) in the CPR are shown in Sec. 3
in case of the primary Yukawa coupling between pions and nucleons. Our
calculations are performed using purely algebraic means within an iterative
technique proposed in that Section. The features that distinguish the
obtained momentum-space quasipotentials from their on-energy-shell
counterparts derived in the second-order Dyson perturbation theory with
Feynman rules are emphasized. Explicit formulae for the quasipotentials
generated by the additional heavier-meson exchange mechanisms are given in
Sec. 4.

\section{Formalism}

\subsection{Introductory definitions}

Our departure point is the Hamiltonian
\begin{equation}
H=H(\stackrel{\circ }{\alpha })=H_0(\stackrel{\circ }{\alpha })+V_0(%
\stackrel{\circ }{\alpha }),  \label{H(BareParticlesWithBareMasses)}
\end{equation}
where the unperturbed Hamiltonian $H_0(\stackrel{\circ }{\alpha })$
and the interaction term $V_0(\stackrel{\circ }{\alpha })$ depend on the
creation and destruction operators of the ``bare'' particles with unphysical
masses and coupling constants. Here, $\stackrel{\circ }{\alpha }$ denotes
the set of all these operators. For example, in case of a spinor (fermion)
field $\psi $ and a neutral pseudoscalar meson field $\phi $ one has to
introduce the operators $\stackrel{\circ }{a}\left( \mathbf{k}\right) $, $%
\stackrel{\circ }{b}\left( \mathbf{p},r\right) $, $\stackrel{\circ }{d}%
\left( \mathbf{p},r\right) $ and their adjoint counterparts, respectively,
for mesons, nucleons and antinucleons. These operators enter in the
expansions

\begin{equation}
\phi \left( \mathbf{x}\right) =\left( 2\pi \right) ^{-3/2}\int d\mathbf{k}%
\left( 2\omega _{\mathbf{k}}^{\circ }\right) ^{-1/2}\left[ \stackrel{\circ }{%
a}(\mathbf{k})+\stackrel{\circ }{a}{}^{\dagger }(-\mathbf{k})\right] \exp (i%
\mathbf{kx}),  \label{A.3}
\end{equation}

\begin{equation}
\pi \left( \mathbf{x}\right) =-i\left( 2\pi \right) ^{-3/2}\int d\mathbf{k}%
\left( \omega _{\mathbf{k}}^{\circ }/2\right) ^{1/2}\left[ \stackrel{\circ }{%
a}(\mathbf{k})-\stackrel{\circ }{a}{}^{\dagger }(-\mathbf{k})\right] \exp (i%
\mathbf{kx}),  \label{1.3}
\end{equation}

\begin{equation}
\psi \left( \mathbf{x}\right) =\left( 2\pi \right) ^{-3/2}\int d\mathbf{p}%
\left( m_0/E_{\mathbf{p}}^{\circ }\right) ^{1/2}\sum_{r,i}\stackrel{\circ }{U%
}_i\left( \mathbf{p},r\right) \stackrel{\circ }{F}_i\left( \mathbf{p}%
,r\right) \exp (i\mathbf{px}).  \label{A.4}
\end{equation}
In the last expression we have introduced the following matrix notations

\begin{equation}
\stackrel{\circ }{U} \left( \mathbf{p},r\right) =
\left(
\begin{matrix}
 \stackrel{\circ}{U}_1  \left( {\bf p},r\right) \cr
  \stackrel{\circ}{U}_2  \left( {\bf p},r\right) \cr
\end{matrix} \right)=
\left(
\begin{matrix}
  \stackrel{\circ }{u} \left( {\bf p},r\right)\cr
  \stackrel{\circ }{v} \left(-{\bf p},r\right)\cr
\end{matrix}\right),
\label{DiracBlockSpinors}
\end{equation}

\begin{equation}
\stackrel{\circ }{F}\left( \mathbf{p},r\right) =
\left(\begin{matrix}
\stackrel{\circ}{F}_1 \left( {\bf p},r\right) \cr
\stackrel{\circ}{F}_2 \left( {\bf p},r\right) \cr
\end{matrix} \right)
=\left(\begin{matrix}
\stackrel{\circ }{b} \left( {\bf p},r\right) \cr
\stackrel{\circ }{d} {}^\dagger \left( -{\bf p},r\right) \cr
\end{matrix}\right).
\label{DiracBlockOperators}
\end{equation}
In other words, the lower index $i$ (the energy-sign index) in $\stackrel{
\circ }{U_i}\left( \mathbf{p},r\right) $and $\stackrel{\circ }{F_i}\left(
\mathbf{p},r\right) $reflects the particle-antiparticle degrees of freedom
in the Dirac formalism. The quantities $\mathbf{k}$, $\mathbf{p}$ and $r$
are the particle momenta and the fermion polarization index. The two Dirac
spinors $\stackrel{\circ }{u}$ and $\stackrel{\circ }{v}$ satisfy the
conventional equations $(\not{p}^{\circ }-m_0)~\stackrel{\circ }{u}~\left(
\mathbf{p} ,r\right) =~0$ and $(\not{p}^{\circ }+m_0)\stackrel{\circ }{v}%
\left( \mathbf{p} ,r\right) =0$ with $\not{p}^{\circ }=E_{\mathbf{p}}^{\circ
} \gamma^0-\mathbf{p} \mbox{{${\bf \gamma}$} }$. The relativistic bare
energies are expressed as $E_{\mathbf{p} }^{\circ }=\sqrt{\mathbf{p}^2+m_0^2}
$ and $\omega _{\mathbf{k}}^{\circ }=\sqrt{\mathbf{k}^2+\mu _0^2}$, where
the unknown values $m_0$ and $\mu _0$ play role of the bare
(nonrenormalized) masses. As in Ref.~\cite{SheShi2001}, we use throughout
this paper the definition of the Dirac matrices, with orthonormalization as
in Bjorken-Drell (Ref.~\cite{BD64}) for $\stackrel{\circ }{u}$ and $%
\stackrel{\circ }{v}$,

\begin{equation}
\stackrel{\circ }{U}_i{}^{\dagger }\left( \mathbf{p},r\right) \stackrel{%
\circ }{U}_j\left( \mathbf{p},r^{\prime }\right) =\frac{E_{\mathbf{p}%
}^{\circ }}{m_0}\delta _{i,j}\delta _{r,r^{\prime }}.  \label{A.11}
\end{equation}

The operators $\stackrel{\circ }{a}(\mathbf{k})$ and $\stackrel{\circ }{a}
{}^{\dagger }(\mathbf{k})$, $\stackrel{\circ }{b}\left( \mathbf{p},r\right) $
and $\stackrel{\circ }{b}^{\dagger }\left( \mathbf{p},r\right) $, $\stackrel{%
\circ }{d}\left( \mathbf{p},r\right) $ and $\stackrel{\circ }{d}^{\dagger
}\left( \mathbf{p} ,r\right)$ satisfy the usual commutation relations

\begin{eqnarray}
\left[ \stackrel{\circ }{a}\left( \mathbf{k}\right) ,\,\stackrel{\circ }{a}
^{\dagger }\left( \mathbf{k}^{\prime }\right) \right] &=&\delta (\mathbf{k}-%
\mathbf{k} ^{\prime }),  \nonumber \\
\left\{ \stackrel{\circ }{b}\left( \mathbf{p},r\right) ,\stackrel{\circ }{b}
^{\dagger }\left( \mathbf{p}^{\prime },r^{\prime }\right) \right\}
&=&\left\{ \stackrel{\circ }{d}\left( \mathbf{p},r\right) ,\stackrel{\circ }{%
d}^{\dagger }\left( \mathbf{p}^{\prime },r^{\prime }\right) \right\} =\delta
_{rr^{\prime }}\delta (\mathbf{p}-\mathbf{p}^{\prime }\mathbf{)}
\label{BareOperatorsCommutationRules}
\end{eqnarray}
which, in particular, yield

\begin{equation}
\left\{ \stackrel{\circ }{F}_i\left( \mathbf{p},r\right) ,\stackrel{\circ }{F%
}_j{}^{\dagger }\left( \mathbf{p}^{\prime },r^{\prime }\right) \right\}
=\delta _{i,j}\delta _{r,r^{\prime }}\delta (\mathbf{p}-\mathbf{p}^{\prime
}).  \label{A.10}
\end{equation}

In the BPR the corresponding unperturbed Hamiltonian is
\begin{equation}
H_0(\stackrel{\circ }{\alpha })=\int d\mathbf{k}\omega _{\mathbf{k}}^{\circ }%
\stackrel{\circ }{a}{}^{\dagger }\left( \mathbf{k}\right) \stackrel{\circ }{a%
}\left( \mathbf{k}\right) +\int d\mathbf{p}E_{\mathbf{p}}^{\circ }\left[
\stackrel{\circ }{b}^{\dagger }\left( \mathbf{p},r\right) \stackrel{\circ }{b%
}\left( \mathbf{p},r\right) +\stackrel{\circ }{d}^{\dagger }\left( \mathbf{p}%
,r\right) \stackrel{\circ }{d}\left( \mathbf{p},r\right) \right] .
\label{Free}
\end{equation}

By definition, the one-bare-fermion and one-bare-meson states $|\mathbf{p}%
,r\rangle ^{\circ }$ and $|\mathbf{k}\rangle ^{\circ }$ are eigenstates of $%
H_0$ with eigenvalues ${E_{\mathbf{p}}^{\circ }}=\sqrt{\mathbf{p}^2+m_0^2}$
and $\omega _{\mathbf{k}}^{\circ }=\sqrt{\mathbf{k}^2+\mu _0^2}$. These
states are built of the bare vacuum $\stackrel{\circ }{\Omega }_0$, $|%
\mathbf{p},r\rangle ^{\circ }=\stackrel{\circ }{b}{}^{\dagger }(\mathbf{p}%
,r)\ \stackrel{\circ }{\Omega }_0$ and $|\mathbf{k}\rangle ^{\circ }=%
\stackrel{\circ }{a}{}^{\dagger }(\mathbf{k})\ \stackrel{\circ }{\Omega }_0$%
, respectively. As usual, the no-bare-particle state $\stackrel{\circ }{%
\Omega }_0$ is destroyed by the operators $\stackrel{\circ }{a}$, $\stackrel{%
\circ }{b}$ and $\stackrel{\circ }{d}$, i. e., $\stackrel{\circ }{a}(\mathbf{%
k})\stackrel{\circ }{\Omega }_0=\stackrel{\circ }{b}(\mathbf{p},r)\stackrel{%
\circ }{\Omega }_0=\stackrel{\circ }{d}(\mathbf{p},r)\stackrel{\circ }{%
\Omega }_0=0$ for ${\forall \mathbf{k},\mathbf{p},r}$.

\subsection{The mass-changing Bogoliubov-type unitary transformation}

Now, let us consider a set $\alpha =(a,a^{\dagger },...)$ of destruction and
creation operators for particles with generic masses $m$ and $\mu $. If $m$
and $\mu $ assume the physical values, the representation refers to ``bare
particles with physical masses'' (see Ref.~\cite{SheShi2001}). By
definition, the operators $\alpha $ enter in the following expansions of the
same fields:

\begin{equation}
\phi \left( \mathbf{x}\right) =\left( 2\pi \right) ^{-3/2}\int d\mathbf{k}%
\left( 2\omega _{\mathbf{k}}\right) ^{-1/2}\left[ a(\mathbf{k})+a^{\dagger
}(-\mathbf{k})\right] \exp (i\mathbf{kx}),  \label{FiMesonFieldDecomposition}
\end{equation}

\begin{equation}
\pi \left( \mathbf{x}\right) =-i\left( 2\pi \right) ^{-3/2}\int d\mathbf{k}%
\left( \omega _{\mathbf{k}}/2\right) ^{1/2}\left[ a(\mathbf{k})-a^{\dagger
}(-\mathbf{k})\right] \exp (i\mathbf{kx}),  \label{PiMesonFieldDecomposition}
\end{equation}

\begin{equation}
\psi \left( \mathbf{x}\right) \mathbf{=}\left( 2\pi \right) ^{-3/2}\int d%
\mathbf{p}\left( m/E_{\mathbf{p}}\right) ^{1/2}\sum_{r,i}U_i\left( \mathbf{p}%
,r\right) F_i\left( \mathbf{p},r\right) \exp (i\mathbf{px}),
\label{NucleonFieldDecomposition}
\end{equation}
where $u\left( \mathbf{p},r\right) $ and $v\left( \mathbf{p},r\right) $ are
the Dirac spinors, which satisfy the equations
\begin{eqnarray*}
(\not{p}-~m)u\left( \mathbf{p},r\right)  &=&0 \\
(\not{p}+m)v\left( \mathbf{p},r\right)  &=&0
\end{eqnarray*}
with $\not{p}=E_{\mathbf{p}}\gamma ^0-\mathbf{p}\mathbf{\gamma }$, $E_{%
\mathbf{p}}=\sqrt{\mathbf{p}^2+m^2}$ and $\omega _{\mathbf{k}}=\sqrt{\mathbf{%
k}^2+\mu ^2}$. The spinor column $U\left( \mathbf{p},r\right) $ is composed
of the spinors $u\left( \mathbf{p},r\right) $ and $v\left( -\mathbf{p}%
,r\right) $ just as the column $\stackrel{\circ }{U}\left( \mathbf{p}%
,r\right) $ of $\stackrel{\circ }{u}\left( \mathbf{p},r\right) $ and $%
\stackrel{\circ }{v}\left( -\mathbf{p},r\right) $. Similarly, the operator
column $F\left( \mathbf{p},r\right) $ is composed of the operators $b\left(
\mathbf{p},r\right) $ and $d^{\dagger }\left( \mathbf{p},r\right) $ just as
the column $\stackrel{\circ }{F}\left( \mathbf{p},r\right) $ of $\stackrel{%
\circ }{b}\left( \mathbf{p},r\right) $ and $\stackrel{\circ }{d}^{\dagger
}\left( \mathbf{p},r\right) $.

Comparing the expressions (\ref{FiMesonFieldDecomposition})-(\ref
{NucleonFieldDecomposition}) and (\ref{A.3})-(\ref{A.4}), we find the links

\begin{equation}
\frac{\stackrel{\circ }{a}(\mathbf{k})+\stackrel{\circ }{a}{}^{\dagger }(-%
\mathbf{k})}{\sqrt{\omega _{\mathbf{k}}^{\circ }}}=\frac{a(\mathbf{k}%
)+a^{\dagger }(-\mathbf{k})}{\sqrt{\omega _{\mathbf{k}}}},\,{\forall \mathbf{%
k},}  \label{A.7}
\end{equation}

\begin{equation}
\left( \stackrel{\circ }{a}(\mathbf{k}) - \stackrel{\circ }{a} {}^{\dagger
}(-\mathbf{k})\right) \sqrt{\omega _{\mathbf{k}}^{\circ }} = \left( a(%
\mathbf{k}) - a^{\dagger }(-\mathbf{k})\right) \sqrt{\omega _{\mathbf{k}}},\,%
{\forall \mathbf{k},}  \label{Similarity ConditionForPiFiels}
\end{equation}

\begin{equation}
\sqrt{\frac{m_0}{E_{\mathbf{p}}^{\circ }}}\sum_{r,i} \stackrel{\circ }{U}_i
\left( \mathbf{p}, r \right) \stackrel{\circ }{F}_i \left( \mathbf{p}, r
\right) =\sqrt{\frac m{E_{\mathbf{p}}}}\sum_{r,i} U_i \left( \mathbf{p},r
\right) F_i \left( \mathbf{p},r \right) ,\,{\forall \mathbf{p.}}  \label{A.8}
\end{equation}

Moreover, operators $\alpha$ are assumed to meet the same commutation rules
as operators $\stackrel{\circ }{\alpha }$ do. Thus, the relations (\ref{A.10}%
) are conserved as well. The orthogonality of $U\left( \mathbf{p}\right) $
follows from Eq. (\ref{A.11}), after the substitution $m_0\rightarrow m$ ($%
E_{\mathbf{p}}^{\circ }\rightarrow E_{\mathbf{p}}$). Since the operators $%
\stackrel{\circ }{\alpha}$ and $\alpha$ meet the same commutation relations,
it is reasonable to look for a similarity (unitary) transformation

\begin{equation}
\stackrel{\circ }{a}\left( \mathbf{k}\right) =Ta\left( \mathbf{k}\right)
T^{\dagger },\,\,\stackrel{\circ }{b}\left( \mathbf{p},r\right) =Tb\left(
\mathbf{p},r\right) T^{\dagger },\,\,\stackrel{\circ }{d}\left( \mathbf{p}%
,r\right) =Td\left( \mathbf{p},r\right) T^{\dagger },  \label{A.1}
\end{equation}
that connects them. We confine ourselves to the form $T=T_{mes}\otimes
T_{ferm}$, where the UT's $T_{mes}$ and $T_{ferm}$ act in mesonic and
fermionic sectors respectively and generate the linear relations:
\begin{equation}
\stackrel{\circ }{a}\left( \mathbf{k}\right) =T_{mes}a\left( \mathbf{k}%
\right) T_{mes}^{\dagger }=c_1(\mathbf{k})a(\mathbf{k})+c_2(\mathbf{k}%
)a^{\dagger }(-\mathbf{k}),  \label{A.12}
\end{equation}
with real functions $c_1(\mathbf{k})$ and $c_2(\mathbf{k})$ and
\begin{equation}
\stackrel{\circ }{F}_i\left( \mathbf{p},r\right) =T_{ferm}F_i\left( \mathbf{p%
},r\right) T_{ferm}^{\dagger }=\sum_{j,r^{\prime }}O_{i,r;j,r^{\prime }}(%
\mathbf{p})F_j\left( \mathbf{p},r^{\prime }\right) ,  \label{A.13}
\end{equation}
where for a given $\mathbf{p}$ the $c$-number coefficients $%
O_{i,r;j,r^{\prime }}(\mathbf{p})$ form a unitary 4$\times $4 matrix $O(%
\mathbf{p})$.

Evidently, the constraint $c_1^2-c_2^2=1$ is necessary to ensure the
commutation $\left[ a\left( \mathbf{k}\right) ,\,a^{\dagger }\left( \mathbf{k%
}^{\prime }\right) \right] =\delta (\mathbf{k}-\mathbf{k}^{\prime })$.

The condition (\ref{A.12}) can be fulfilled if one introduces the following
ansatz \footnote{%
Such a form has been prompted by M.Shirokov during a tentative investigation
carried out by him and one of us (A.S.)}
\begin{equation}
T_{mes}=\exp \left[ -\frac 12\int d\mathbf{k}\chi _k\left( a^{\dagger
}\left( \mathbf{k}\right) a^{\dagger }\left( -\mathbf{k}\right) -a\left(
\mathbf{k}\right) a\left( -\mathbf{k}\right) \right) \right] ,  \label{A.14}
\end{equation}
where $\chi _k$ is some real function of $k=\left| \mathbf{k}\right| $.

In fact, then we get
\begin{equation}
\stackrel{\circ }{a}\left( \mathbf{k}\right) =\cosh \chi _ka(\mathbf{k}%
)+\sinh \chi _ka^{\dagger }(-\mathbf{k}),  \label{A.15}
\end{equation}
and Eq. (\ref{A.7}) holds if
\begin{equation}
\exp \chi _k=\sqrt{\frac{\omega _{\mathbf{k}}^{\circ }}{\omega _{\mathbf{k}}}%
},  \label{A.16}
\end{equation}
\textit{i.e.},
\begin{eqnarray*}
\cosh \chi _k &=&\frac 12\left[ \sqrt{\frac{\omega _{\mathbf{k}}^{\circ }}{%
\omega _{\mathbf{k}}}}+\sqrt{\frac{\omega _{\mathbf{k}}}{\omega _{\mathbf{k}%
}^{\circ }}}\right] , \\
\sinh \chi _k &=&\frac 12\left[ \sqrt{\frac{\omega _{\mathbf{k}}^{\circ }}{%
\omega _{\mathbf{k}}}}-\sqrt{\frac{\omega _{\mathbf{k}}}{\omega _{\mathbf{k}%
}^{\circ }}}\right] .
\end{eqnarray*}
It can be readily verified that with the choice (\ref{A.16}) the condition (%
\ref{Similarity ConditionForPiFiels}) is automatically satisfied.

In regards to the fermionic sector, starting from Eq. (\ref{A.8}) and
omitting the discrete indices, we find
\begin{equation}
\sqrt{\frac{m_0}{E_{\mathbf{p}}^{\circ }}} \stackrel{\circ }{U}\left(
\mathbf{p} \right) O(\mathbf{p}) = \sqrt{\frac m{E_{\mathbf{p}}}} U\left(
\mathbf{p} \right) .  \label{A.17}
\end{equation}

Further, the orthogonality for the spinors $\stackrel{\circ }{U}$ enables us
to write
\begin{eqnarray}
O(\mathbf{p}) &=&\sqrt{\frac{m\,m_0}{E_{\mathbf{p}}E_{\mathbf{p}}^{\circ }}}%
\stackrel{\circ }{U}{}_i^{\dagger }\left( \mathbf{p},r\right) U_j\left(
\mathbf{p},r^{\prime }\right)  \nonumber \\
&=&\sqrt{\frac{m\,m_0}{E_{\mathbf{p}}E_{\mathbf{p}}^{\circ }}}
\left(\begin{matrix}
\stackrel{\circ }{u} {}^{\dagger }\left( {\bf p},r \right) u\left( {\bf
p},r^{\prime }\right) &
\stackrel{\circ }{u} {}^{\dagger }\left( {\bf p},r
\right) v\left( -{\bf p},r^{\prime }\right) \cr \stackrel{\circ }{v}
{}^{\dagger }\left( -{\bf p},r \right) u\left( {\bf p},r^{\prime }\right) &
\stackrel{\circ }{v} {}^{\dagger }\left( -{\bf p},r \right) v\left( -{\bf
p},r^{\prime }\right) \cr \end{matrix}\right).  \label{A.18}
\end{eqnarray}

The diagonal elements of the matrix (\ref{A.18}) turn out to be equal to
each other:

\[
\stackrel{\circ }{u}{}^{\dagger }\left( \mathbf{p},r\right) u\left( \mathbf{p%
},r^{\prime }\right) =\stackrel{\circ }{v}{}^{\dagger }\left( -\mathbf{p}%
,r\right) v\left( -\mathbf{p},r^{\prime }\right) =
\]
\[
\frac 12\sqrt{\frac{(E_{\mathbf{p}}+m)(E_{\mathbf{p}}^{\circ }+m_0)}{mm_0}}%
\left\{ 1+\frac{\mathbf{p}^2}{(E_{\mathbf{p}}+m)(E_{\mathbf{p}}^{\circ }+m_0)%
}\right\} \delta _{r,r^{\prime }},
\]
while the off-diagonal elements differ from each other only by sign:

\[
\stackrel{\circ }{u}{}^{\dagger }\left( \mathbf{p},r\right) v\left( -\mathbf{%
p},r^{\prime }\right) =-\stackrel{\circ }{v}{}^{\dagger }\left( -\mathbf{p}%
,r\right) u\left( \mathbf{p},r^{\prime }\right) =
\]
\[
\frac 12\sqrt{\frac{(E_{\mathbf{p}}+m)(E_{\mathbf{p}}^{\circ }+m_0)}{mm_0}}%
\left\{ \frac 1{E_{\mathbf{p}}^{\circ }+m_0}-\frac 1{E_{\mathbf{p}%
}+m}\right\} u^{\dagger }\left( 0,r\right) \mathbf{p}\mathbf{\gamma }v\left(
0,r^{\prime }\right) ,
\]
where $u^{\dagger }\left( 0,r\right) $ and $v\left( 0,r^{\prime }\right) $
are the Dirac spinors in the rest frame and the usual Dirac matrix $\mathbf{%
\gamma }$ is expressed as $\left(
\begin{array}{ll}
0 & \mathbf{\sigma } \\
-\mathbf{\sigma } & 0
\end{array}
\right) $.

This allows to rewrite the unitary transformation for fermions in the
following way
\begin{equation}
O(\mathbf{p})=A\left( p\right) I+B\left( p\right) \mathbf{p\Gamma },
\label{A.20}
\end{equation}
where, along with the unit matrix $I$, we have introduced the 4$\times $4
matrix
\[
\mathbf{\Gamma }=\left[
\begin{array}{ll}
0 & \mathbf{\sigma } \\
-\mathbf{\sigma } & 0
\end{array}
\right]
\]
in the space of energy-sign and polarization indices.

The functions $A(p)$ and $B(p)$ are equal to:

\[
A(p)=\frac 12\sqrt{\frac{(E_{\mathbf{p}}+m)(E_{\mathbf{p}}^{\circ }+m_0)}{E_{%
\mathbf{p}}E_{\mathbf{p}}^{\circ }}}\left\{ 1+\frac{\mathbf{p}^2}{(E_{%
\mathbf{p}}+m)(E_{\mathbf{p}}^{\circ }+m_0)}\right\} ,
\]
\[
B(p)=\frac 12\sqrt{\frac{(E_{\mathbf{p}}+m)(E_{\mathbf{p}}^{\circ }+m_0)}{E_{%
\mathbf{p}}E_{\mathbf{p}}^{\circ }}}\left\{ \frac 1{E_{\mathbf{p}}^{\circ
}+m_0}-\frac 1{E_{\mathbf{p}}+m}\right\} ,
\]
and unitarity ($O^{\dagger }O=OO^{\dagger }=I$) leads to the relation
\begin{equation}
A^2\left( p\right) +\mathbf{p}^2B^2\left( p\right) =1.  \label{A.21}
\end{equation}

Furthermore, as in the meson case, one could start from the ansatz
\begin{equation}
T_{ferm}=\exp \left[ \int d\mathbf{p}F^{\dagger }\left( \mathbf{p}\right)
h\left( \mathbf{p}\right) F\left( \mathbf{p}\right) \right] ,  \label{A.22}
\end{equation}
where $h\left( \mathbf{p}\right) $ is an antihermitian 4$\times $4 matrix.
With such a form we get
\begin{equation}
O\left( \mathbf{p}\right) =\exp \left[ h\left( \mathbf{p}\right) \right] ,
\label{A.23}
\end{equation}
assuming that $h\left( \mathbf{p}\right) =\mathbf{G}\left( \mathbf{p}\right)
\mathbf{\Gamma }$ and taking into account that
\[
\exp \left[ \mathbf{G}\left( \mathbf{p}\right) \mathbf{\Gamma }\right] =\cos
\left( \left\| \mathbf{G}\left( \mathbf{p}\right) \right\| \right) +\frac{%
\mathbf{G}\left( \mathbf{p}\right) \mathbf{\Gamma }}{\left\| \mathbf{G}%
\left( \mathbf{p}\right) \right\| }\sin \left( \left\| \mathbf{G}\left(
\mathbf{p}\right) \right\| \right) .
\]
Here $\left\| \mathbf{G}\left( \mathbf{p}\right) \right\| $ denotes the
length of the vector $\mathbf{G}\left( \mathbf{p}\right) $.

Comparing this with Eq. (\ref{A.20}), we can write $\mathbf{G}\left( \mathbf{%
p}\right) =\mathbf{p}Q\left( p\right) $ where the real function $Q\left(
p\right) $ is determined by
\begin{equation}
pB\left( p\right) =\sin \left[ pQ\left( p\right) \right] .  \label{A.24}
\end{equation}

After we have specified the transformation in question, the free Hamiltonian
can be expressed in terms of the operators $\alpha$,

\begin{equation}
H_0(\stackrel{\circ }{\alpha })=H_F(\alpha )+M_{ren,\,mes}\left( \alpha
\right) +M_{ren,\ ferm}\left( \alpha \right) ,  \label{HOandHF}
\end{equation}
where

\begin{equation}
H_F(\alpha )=\int d\mathbf{k}\omega _{\mathbf{k}}a^{\dagger }\left( \mathbf{k%
}\right) a\left( \mathbf{k}\right) +\int d\mathbf{p}E_{\mathbf{p}}\left[
b^{\dagger }\left( \mathbf{p},r\right) b\left( \mathbf{p},r\right)
+d^{\dagger }\left( \mathbf{p},r\right) d\left( \mathbf{p},r\right) \right] ,
\label{H_F_InParticleForm}
\end{equation}

\begin{equation}
M_{ren,\, mes}\left( \alpha \right) ={\frac{\mu_0^2 - \mu^2}{{4}}}\int \ {%
\frac{d\mathbf{k}}{{\omega _{\mathbf{k}}}}}\left[ {a}^{\dagger }\left(
\mathbf{k}\right) a\left( \mathbf{k}\right) +a\left( \mathbf{k}\right)
a\left( -\mathbf{k}\right) +H.c. \right],
\label{MesonMassCountertermInParticleForm}
\end{equation}

\begin{equation}
M_{ren,\ ferm}\left( \alpha \right) =m(m_0-m)\int \frac{d\mathbf{p}}{E_{%
\mathbf{p}}}\sum_{r,r^{\prime },i,j}F_i^{\dagger }\left( \mathbf{p}%
,r^{\prime }\right) \,M_{i,j}\left( \mathbf{p},r^{\prime };\mathbf{p}%
,r\right) F_j\left( \mathbf{p},r\right) ,
\label{FermionMassCountertermInParticleForm}
\end{equation}
with $M_{i,j}\left( \mathbf{p},r^{\prime };\mathbf{p},r\right) =\bar{U}%
_i\left( \mathbf{p},r^{\prime }\right) U_j\left( \mathbf{p},r\right) $,
where $\bar{U}_i\left( \mathbf{p},r^{\prime }\right)$ $(i=1,2)$ is the
element of the row $\left[\bar{u}\left( \mathbf{p},r\right),\bar{v}\left( -%
\mathbf{p},r\right) \right]$. The use of the subscript ``ren'' associated
with the term ``renormalization'' may be justified \textit{a posteriori}
(see Subsect. 2.3).

The linear expressions (\ref{A.12}) and (\ref{A.13}) perform the transition
from the initial particle representation with primary (bare) masses to the
auxiliary particle representation with other (\textit{e.g.}, physical)
masses. In the latter we have a new vacuum state $\Omega _0=T^{\dagger }%
\stackrel{\circ }{\Omega }_0$ that is destroyed by the new operators $a,b,d$
just like the operators $\stackrel{\circ }{a},\stackrel{\circ }{b},\stackrel{%
\circ }{d}$ do the same with the primary vacuum $\stackrel{\circ }{\Omega }%
_0 $. It is noteworthy to observe that the r.h.s. of Eq. (\ref{HOandHF})
involves the terms that do not conserve the number of bare particles with
new masses. Simultaneously, we have constructed one of possible unitarily
equivalent Fock representations for the system of noninteracting bosons
(mesons) and fermions (nucleons and antinucleons). In order to differentiate
this representation from the primary representation $\stackrel{\circ }{%
\alpha }$ we will refer to it as the $\alpha $ representation. Its vacuum
state $\Omega _0$ , the one-particle states ${a}^{\dagger }\Omega _0$, ${b}%
^{\dagger }\Omega _0$ and ${d}^{\dagger }\Omega _0$ together with other
many-meson and many-fermion $H_F$ eigenstates form the basis of the Fock
space, spanned by the primary basis $\stackrel{\circ }{\Omega }_0$, $%
\stackrel{\circ }{a}{}^{\dagger }\stackrel{\circ }{\Omega }_0$, $\stackrel{%
\circ }{b}{}^{\dagger }\stackrel{\circ }{\Omega }_0$, $\stackrel{\circ }{d}%
{}^{\dagger }\stackrel{\circ }{\Omega }_0$, etc.

At this point, one should note that the transition (\ref{A.12}) has much in
common with the so-called \emph{cosh-sinh} transformation for pair of boson
operators, introduced by Bogoliubov \cite{Bogoliubov1947} in the theory of
the $^3$He superfluidity. However, such a resemblance is rather mathematical
since the Bogoliubov transformation and the UT $T_{mes}$ have different
scopes. As a matter of facts, the Bogoliubov transformation is related to
the Hamiltonian of weakly interacting bosons to reduce it to diagonal form
in the representation of quasiparticles (for a clear and simple discussion
on this subject we refer to Chapter 30 of Ref.~\cite{Kaempffer1965}). In
that respect, the argument $\chi _k$ is determined there with the help of a
completely different physical condition. Instead, in the case of the
transformation (\ref{A.12}) we deal with free bosons and move to the
opposite direction: from the diagonal form of Eq.~(\ref{Free}), to the form (%
\ref{HOandHF}) in which the number of bosons with new mass $\mu $ is not
conserved.

Similarly, the transition (\ref{A.13}) with the matrix $O$ determined by (%
\ref{A.20}) is an analog of the Bogoliubov \emph{u-v} transformation in the
theory of superconductivity (see, \emph{e.g.}, Ref.~\cite
{BogoliubovTolmachevShirkov1959}). The latter is a particular case of the
general canonical transformation\footnote{%
Here the term canonical applies to the transformation which does not violate
the commutation relations for the fermions in question.} for operators of
fermion pairs, after Koppe and M\"{u}hlschlegel~\cite{KoppeMuhlschlegel1968}.

\subsection{Unitary clothing transformations and mass renormalization}

We discuss now some key points of the clothing procedure in QFT (more
details can be found in Refs.~\cite{SheShi2001} and \cite{SheShi2000}). To
be specific, we consider the model of meson-nucleon pseudo scalar (PS)
coupling, in which $V_0=V_0(\stackrel{\circ }{\alpha })=ig_0\bar{\psi}\gamma
_5\psi \phi =ig_0$ $Y(\stackrel{\circ }{\alpha };m_0,\mu _0)$. We refer to
models of this type as Yukawa models. At this point, one should stress that
the described transition to the $\alpha $-representation does not change the
primary interaction $V_0=ig_0Y(\alpha ;m,\mu )$ with $Y(\alpha ;m,\mu )=Y(%
\stackrel{\circ }{\alpha };m_0,\mu _0)$.

By using the decomposition (\ref{HOandHF}), we divide the total Hamiltonian $%
H(\alpha )$ into the new free part $H_F(\alpha )$ and the new interaction
term $H_I(\alpha )$,

\begin{eqnarray}
H&=&H( \alpha )=H_{F}( \alpha )+H_{I}( \alpha ),  \nonumber \\
H_{I}( \alpha )&=&V( \alpha ) + M_{ren,mes}(\alpha) + M_{ren,ferm}(\alpha) +
V_{ren} (\alpha),  \label{H(BareParticlesWithPhysicalMasses)}
\end{eqnarray}
where the operators $M_{ren}(\alpha )$ are considered as the mass
counterterms. Here we assume that the operator $V=ig\bar{\psi}\gamma _5\psi
\phi$ depends on the ``physical'' value of the strength parameter $g$, so
that a vertex counterterm appears in the interaction term $H_I(\alpha )$, $%
V_{ren}(\alpha )\equiv V_0(\alpha )-V(\alpha )$. One could proceed further
in the discussion by introducing properly regularized interaction $V=V_{reg}$
with the cutoff vertex functions, but we will postpone such an extension to
another occasion.

In order to motivate our transition to the CPR, we recall again that the
vacuum state $\Omega _0$, with no bare particles with physical masses, and
the one-particle states ${a}^{\dagger }(\mathbf{k})\Omega _0$, ${b}^{\dagger
}(\mathbf{p},r)\Omega _0$ and ${d}^{\dagger }(\mathbf{p},r)\Omega _0$ are $%
H_F$ eigenvectors, but are not eigenstates of the total Hamiltonian $H$.
This is clearly seen in the current Yukawa model, where the interaction
operator in the $\alpha $-representation looks like
\begin{eqnarray}
V_0(\alpha ) &=&\int d\mathbf{k}\,\hat{V}^{\mathbf{k}}\,a(\mathbf{k})+H.c.,
\nonumber \\
\hat{V}^{\mathbf{k}} &=&\int d\mathbf{p}^{\prime }d\mathbf{p}%
\sum_{r,r^{\prime },i,j}F_i^{\dagger }\left( \mathbf{p}^{\prime },r^{\prime
}\right) \,V_{i,j}^{\mathbf{k}}\left( \mathbf{p}^{\prime },r^{\prime };%
\mathbf{p},r\right) F_j\left( \mathbf{p},r\right) .
\label{V_InBareOperatorForm}
\end{eqnarray}
In the above equation, we have introduced the 2$\times $2 c-number matrices
(cfr. App. A of Ref.~\cite{SheShi2001}),
\begin{equation}
V_{i,j}^{\mathbf{k}}\left( \mathbf{p}^{\prime },r^{\prime };\mathbf{p}%
,r\right) =\frac{ig_0}{\left( 2\pi \right) ^{3/2}}\frac m{\sqrt{2\omega _{%
\mathbf{k}}E_{\mathbf{p}^{\prime }}E_{\mathbf{p}}}}\delta \left( \mathbf{p}+%
\mathbf{k}-\mathbf{p}^{\prime }\right) \bar{U}_i\left( \mathbf{p}^{\prime
},r^{\prime }\right) \gamma _5U_j\left( \mathbf{p},r\right) .
\label{V_SpinorPart}
\end{equation}

This interaction contains, for instance, terms responsible for the
``non-diagonal'' transitions, such as $\Omega _0\longrightarrow {a}^{\dagger
}{b}^{\dagger }{d}^{\dagger }\Omega _0$ and ${b}^{\dagger }\Omega
_0\longrightarrow {a}^{\dagger }{b}^{\dagger }{bb}^{\dagger }\Omega _0$.
Together with the ${a}^{\dagger }{a}^{\dagger }$ and ${b}^{\dagger }{d}%
^{\dagger }$ contributions to $M_{ren,mes}$ and $M_{ren,ferm}$ they prevent
the aforementioned vectors to be eigenvectors of the total Hamiltonian $H$.

To overcome this problem, the CPR introduces another representation of the
total field Hamiltonian,
\begin{equation}
H=K(\alpha _c)=K_F(\alpha _c)+K_I(\alpha _c),  \label{H(ClothedParticles)}
\end{equation}
where the decomposition into a free part $K_F(\alpha _c)$ and an interaction
part $K_I(\alpha _c)$ depend on newly defined destruction and creation
operators $\alpha _c$,
\begin{equation}
a_c\left( \mathbf{k}\right) \left( a_c^{\dagger }\left( \mathbf{k}\right)
\right) ,\ b_c\left( \mathbf{p},r\right) \left( b_c^{\dagger }\left( \mathbf{%
p},r\right) \right) ,\ d_c\left( \mathbf{p},r\right) \left( d_c^{\dagger
}\left( \mathbf{p},r\right) \right) ,{\forall \mathbf{k},\mathbf{p},r}.
\label{ClothedParticleOperators}
\end{equation}
These are called the clothed particle operators\footnote{%
As noted in Introduction, we employ the term "clothed" instead its
synonymous "dressed", since the latter is sometimes used in a sense which
differs from that defined below by the items i) -- iv)}. By definition, they
have the following properties:

i) The physical vacuum (the $H$ lowest eigenstate) must coincide with a new
no--particle state $\Omega $, \emph{i.e}., the state that obeys the
equations
\begin{equation}
a_c\left( \mathbf{k}\right) \left| \Omega \right\rangle =b_c\left( \mathbf{p}
,r\right) \left| \Omega \right\rangle =d_c\left( \mathbf{p},r\right) \left|
\Omega \right\rangle =0,\,\forall \mathbf{k,p,r}
\label{ClothedOperatorDefinition}
\end{equation}
\[
\left\langle \Omega \mathbf{|}\Omega \right\rangle =1.
\]

ii) New one-clothed-particle states $| \mathbf{k} \rangle_c \equiv
a_c^{\dagger }\left( \mathbf{k}\right) \Omega $ etc. are the eigenvectors of
both $K_F$ and $K$.

iii) The spectrum of indices that enumerate the new operators must be the
same as that for the bare ones.

iv) The new operators satisfy the same commutation rules as do their bare
counterparts. For instance,
\begin{equation}
\left[ a_{c}\left( \mathbf{k}\right) ,a_{c}^{\dagger }\left( \mathbf{k}%
^{\prime }\right) \right] =\delta (\mathbf{k}-\mathbf{k}^{\prime }),\left\{
b_{c}\left( \mathbf{p},r\right) ,b_{c}^{\dagger }\left( \mathbf{p}^{\prime
},r^{\prime }\right) \right\} =\left\{ d_{c}\left( \mathbf{p},r\right)
,d_{c}^{\dagger }\left( \mathbf{p}^{\prime },r^{\prime }\right) \right\}
=\delta _{rr^{\prime }}\delta (\mathbf{p}-\mathbf{p}^{\prime }\mathbf{).}
\label{CommutationRules}
\end{equation}

To be more specific, the property ii) implies that if the clothed meson
state $| \mathbf{k}\rangle _c$ belongs to the eigenvalue $\omega _{\mathbf{k}%
}=\sqrt{\mathbf{k} ^2+\mu ^2}$ of the operator $K_F$, then $K|\mathbf{k}%
\rangle _c=\omega _{\mathbf{k} }|\mathbf{k}\rangle _c$. In other words, the
operator $K_I$ to be found must possess the property:
\begin{equation}
K_I|\mathbf{k}\rangle _c=0.  \label{KIProperty}
\end{equation}
The same is valid for the clothed fermion states. This property defines
an important distinctive feature of the CPR.

Now, when finding the operators $\alpha_c$ as functions of $\alpha$, in
order to meet the property iii) we suppose $\alpha_c = W^\dagger \alpha W$,
where $W = W(\alpha) = W(\alpha_c) = \mathrm{exp} R(\alpha_c)$ is a UT $%
(WW^\dagger = W^\dagger W =1)$. Before constructing its generator $%
R(\alpha_c ) = - R^\dagger (\alpha_c )$, let us rewrite the total
Hamiltonian as

\begin{eqnarray}
H&=&H(\alpha)=H\left( W\left( \alpha_c \right) \alpha_c W^{\dagger
}\left(\alpha_c \right) \right)=W\left( \alpha_c \right) H\left( \alpha_c
\right)W^{\dagger }\left( \alpha_c \right) =K\left( \alpha_c \right)
\nonumber \\
&=&H_{F}\left( \alpha_c \right) +H_{I}\left( \alpha_c \right)
+\left[R,H_{F}\right] +\left[ R,H_{I}\right]  \nonumber \\
&+&{\frac{1}{2}}[R,\left[ R,H_{F}\right]]+{\frac{1}{2}}\left[ R,\left[
R,H_{I}\right] \right] +...
\label{FinalToInitialHamiltonianExplicitRelation}
\end{eqnarray}

It is important to realize that the operator $K\left( \alpha_c \right)$ is
the same Hamiltonian as $H(\alpha)$ but it has another dependence on its
argument $\alpha_c$ as compared to $H(\alpha)$ because it refers to a
different representation. Also one should note that the new free part $K_F
\equiv K_F (\alpha_c) \ne H_F (\alpha) \equiv H_F $, but $K_F (\alpha_c)=H_F
(\alpha_c)$. Hence, comparing Eqs. (\ref{H(ClothedParticles)}) and (\ref
{FinalToInitialHamiltonianExplicitRelation}) we see that $K_{I}\left(
\alpha_c \right) = K\left( \alpha_c \right) - H_{F}\left( \alpha_c \right) $.

Eq. (\ref{FinalToInitialHamiltonianExplicitRelation}) gives a practical
recipe for the $K\left( \alpha_c \right) $ calculation: at the beginning one
replaces $\alpha$ by $\alpha_c $ in the initial expression $H\left( \alpha
\right) $ and then calculates $W\left( \alpha_c \right) H\left( \alpha_c
\right)W^{\dagger }\left( \alpha_c \right) $ using Eqs. (\ref
{FinalToInitialHamiltonianExplicitRelation}) and (\ref{CommutationRules}).
The above transition $H\left( \alpha \right) \rightarrow H\left( \alpha_c
\right) $ generates a new operator $H\left( \alpha_c \right) $ as compared
to $H\left( \alpha \right) $, but Eq. (\ref
{FinalToInitialHamiltonianExplicitRelation}) show that $W\left( \alpha_c
\right)H\left( \alpha_c \right) W^{\dagger }\left( \alpha_c \right) $ turns
out to be equal to the original total Hamiltonian.

Further, to meet the requirements i) and ii), the r.h.s. of Eq. (\ref
{FinalToInitialHamiltonianExplicitRelation}) must not contain some
undesirable terms that prevent the no--clothed--particle state $\Omega $ and
one-clothed--particle states to be eigenstates of the total Hamiltonian.
Such terms (we call them bad as in Ref. \cite{SheShi2000}) enter in the
operator $V(\alpha _c)$ that is derived from $V(\alpha )$ by means of the
replacement $\alpha \rightarrow \alpha _c$. By definition, these terms,
taken together with their H.c. counterparts, do not destroy the physical
vacuum and simultaneously the one-clothed-particle states.

In case of the Yukawa model all terms in the r.h.s. of Eq.~(\ref
{V_InBareOperatorForm}) are bad, and to eliminate them from $K
\left(\alpha_c \right) $ we choose an $R$ that satisfies the relation
\begin{equation}
V+\left[ R,H_F\right] =0\ .  \label{EquationForR^1}
\end{equation}

It turns out that a solution of this equation can be represented as \cite
{SheShi2001}

\begin{equation}
R=-i\lim_{\varepsilon \rightarrow 0+}\int_0^\infty V(t)\mathrm{e}%
^{-\varepsilon t}\mathrm{d}t\ ,  \label{R_InIntegralOperatorForm}
\end{equation}
where $V(t)=\mathrm{exp}(iH_{\mathrm{F}}t)V\mathrm{exp}(-iH_{\mathrm{F}}t)$
is the interaction operator in the Dirac picture.

According to \cite{SheShi2001}, the corresponding operator $R$ can be
expressed as $R=\mathcal{R}-\mathcal{R}^{\dagger }$ with
\begin{eqnarray*}
\mathcal{R}=\int d\mathbf{k}\hat{R}_c^{\mathbf{k}}\,a_c(\mathbf{k}),
\end{eqnarray*}
\begin{equation}
\hat{R}_c^{\mathbf{k}}=\int d\mathbf{p}^{\prime }d\mathbf{p}%
\sum_{r,r^{\prime },i,j}F_{c,i}^{\dagger }\left( \mathbf{p}^{\prime
},r^{\prime }\right) \,R_{i,j}^{\mathbf{k}}\left( \mathbf{p}^{\prime
},r^{\prime };\mathbf{p},r\right) F_{c,j}\left( \mathbf{p},r\right) .
\label{R_InClothedOperatorForm}
\end{equation}
Here, unlike the fermion operators $F$ and $F^{\dagger }$ in Eq. (\ref
{V_InBareOperatorForm}) the operator column $F_c$ and row $F_c^{\dagger }$
are composed of the clothed nucleon and antinucleon operators. As shown in
\cite{SheShi2001}, the c-number 2$\times $2 matrix $R^{\mathbf{k}}$ is
determined by
\begin{equation}
R_{i,j}^{\mathbf{k}}\left( \mathbf{{p}\,^{\prime },}r\mathbf{^{\prime }};%
\mathbf{{p}\,,}r\right) =V_{i,j}^{\mathbf{k}}\left( \mathbf{{p}\,^{\prime },}%
r\mathbf{^{\prime }};\mathbf{{p}\,,}r\right) /\left[ \left( -1\right)
^{i-1}E_{\mathbf{p}^{\prime }}-\left( -1\right) ^{j-1}E_{\mathbf{p}}-\omega
_{\mathbf{k}}\right] ,\,\left( i,j=1,2\right) .  \label{R_SpinorPart}
\end{equation}

One should note that the solution of Eq. (\ref{EquationForR^1}) exists if $%
\mu <2m$. This condition has a clear physical meaning, in that, if $\mu >2m$%
, the meson can decay into the $f\bar{f}$--pair making one--meson states not
stable, namely, they cannot be $H$ eigenvectors.

Once $[R,H_F]=-V$, Eq. (\ref{FinalToInitialHamiltonianExplicitRelation}) can
be rewritten as follows
\begin{equation}
K\left( \alpha _c\right) =H_F\left( \alpha _c\right) +M_{ren}\left( \alpha
_c\right) +V_{ren}\left( \alpha _c\right) +{\frac 12}\left[ R,V\right]
+\left[ R,M_{ren}\right] +{\frac 13}\left[ R,\left[ R,V\right] \right] +\
...\ .  \label{K_AfterFirstClothing}
\end{equation}
Doing so, we have removed from $K\left( \alpha _c\right) $ all the bad terms
of the $g^1$--order and left the contributions up to the $g^3$--order terms
assuming that $M_{ren}\sim O(g^2)$ and $V_{ren}\sim O(g^3)$.

However, the r.h.s. of Eq. (\ref{K_AfterFirstClothing}) embodies other bad
terms of the $g^2$- and higher orders. The commutator $\left[ R,V\right] $
involves the terms $g^2b_{{c}}^{\dagger }d_{{c}}^{\dagger }a_{{c}}^{\dagger
}a_{{c} }^{\dagger }$ , which do not destroy the physical vacuum $\Omega $ .
In addition, we find in $\left[R,V\right] $ the terms $g^2b_{{c}}^{\dagger
}d_{{\ c}}^{\dagger } a_{{c} }^{\dagger }a_{{c}}$, which neither destroy $a_{%
{c} }^{\dagger }\Omega $, nor retain it with a multiplicative factor. These
and similar bad terms can be eliminated with one more transformation, in a
way analogous to the described above.

The commutator $[R,V]$ also contains the meson and fermion two-operator
terms $a^\dagger a$, $a a$, $b^\dagger b$, $b^\dagger d^\dagger$ and so on,
whose structure repeats the structure of the expansions~(\ref
{MesonMassCountertermInParticleForm}) and (\ref
{FermionMassCountertermInParticleForm}). Not all of them are bad (for
instance, $a^\dagger a$). It is required that the ``diagonal''
(particle-conserving number) species of the $a^\dagger a$-, $b^\dagger b$-,
and $d^\dagger d$- types cancel the corresponding contributions to the mass
counterterms $M_{ren,\, mes}\left( \alpha_c \right)$ and $M_{ren,\,
ferm}\left( \alpha_c \right)$. Note that it is sufficient to evaluate the
mass shifts $m-m_0$ and $\mu^2-\mu_0^2$ in the $g^2$-order; since the same
operator structure will appear in higher orders in the coupling constant we
can extend this requirement to determine these mass shifts order by order.

The first results in this direction have been obtained in \cite{SheShi2001},
\cite{KordaShebekoPhys.Rev.2004} for interacting pions and nucleons in the
Yukawa model with the PS coupling. For example, according to \cite
{SheShi2001}, the meson mass shift in the $g^2$ -order is equal to
\begin{equation}
\delta \mu ^2\equiv \mu _0^2-\mu ^2=\frac{2g{^2}}{\left( 2\pi \right) ^3}%
\int \frac{d\mathbf{p}}{E_{\mathbf{p}}}\left\{ 1+\frac{{\mu }^4}{4\left(
pk\right) ^2-{\mu }^4}\right\} .  \label{MesonMassShiftInG^2}
\end{equation}
The nucleon mass shift evaluated in \cite{KordaShebekoPhys.Rev.2004} in the
same order can be written as
\begin{eqnarray}
\delta m &\equiv &m_0-m=\frac{g^2}{4m(2\pi )^3}\left[ I_1(p)+I_2(p)\right] ,
\nonumber \\
I_1(p) &=&\int \frac{d\mathbf{k}}{\omega _{\mathbf{k}}}pk\left\{ \frac{{\ }1%
}{{\ }\mu ^2-2pk}-\frac{{\ }1}{\ \mu ^2+2pk}\right\} ,  \nonumber \\
I_2(p) &=&\int \frac{d\mathbf{q}}{E_{\mathbf{q}}}\left\{ \frac{{\ }m^2-pq}{\
2\left[ m^2-pq\right] -\mu ^2}+\frac{m^2+pq}{{\ }2\left[ m^2+pq\right] -\mu
^2}\right\} .  \label{NucleonMassShiftInG^2}
\end{eqnarray}
Here we have adopted the four-vector notation, namely $p=\left( E_{\mathbf{p}%
},\mathbf{p}\right) $, $k=\left( \omega _{\mathbf{k}},\mathbf{k}\right) $
and $q=\left( E_{\mathbf{q}},\mathbf{q}\right) $.

We observe that, being expressed through the explicitly covariant integrals
in the r.h.s of Eqs. (\ref{MesonMassShiftInG^2}) and (\ref
{NucleonMassShiftInG^2}), these quantities do not depend on the particle
momenta. It turns out \cite{KordaShebekoPhys.Rev.2004} that these integrals
coincide with the corresponding one-loop Feynman integrals. In this
connection, we observe that recent calculations in \cite
{KrugerGlockle.Phys.Rev.C60.024004.1999} have been carried out using Okubo's
idea with a toy boson-fermion interaction model. The result obtained there
for the fermion mass shift reproduces the corresponding expression deduced
via Feynman-diagram technique.

Up to now, the way the mass parameters have been introduced appears, to a
great extent, quite artificial. However, once $m$ and $\mu$ are identified
with the observed physical masses, then, the eigenvalue ${E_{\mathbf{p}}}$
in the equation $H$ $|\mathbf{p},r\rangle =$ ${E_{\mathbf{p}}}|\mathbf{p}%
,r\rangle$, with ${E_{\mathbf{p}}}=\sqrt{\mathbf{p}^2+m^2}$, becomes the
fermion energy. And analogously, one can discuss the appearance in the CPR
of the meson physical mass $\mu$, which is different from the trial mass $%
\mu _0$. This represents, in our opinion, a very natural way to introduce
(renormalized) masses in the system.

Finally, we would like to stress that the mass renormalization is
implemented here in concomitance with the construction of relativistic
interactions involved in operator $K_I$ and describing the physical
exchange-type processes between clothed particles. In the following, we will
give a reasonable approximation to the operator $K_I$ that must meet the
condition (\ref{KIProperty}).

\section{The construction of relativistic interactions via unitary clothing
transformations}

Before deriving analytic expressions for the interactions in the CPR one
should emphasize that the corresponding operators stem primarily from the
commutators $\left[ R,V\right] $, $\left[ R,\left[ R,V\right] \right] $,
etc. in the r.h.s. of Eq.~(\ref{K_AfterFirstClothing}). 
It is convenient to classify such operators by the numbers of creation
and annihilation operators that appear in the normal ordered product.
According to this classification~\cite{SheShi2001}, an operator is 
of class $[m,n]$ if its normal
product is made of $m$ creation and $n$ annihilation operators. In fact, 
after normal ordering of the creation (destruction) operators $\alpha _c$ 
one can separate out the ``good'' four-operator contributions 
of the $g^2$-order $V_{good}^{(2)}\left( \alpha _c\right) \sim 
a_{{c}}^{\dagger }b_{{c}}^{\dagger }a_{{c}}b_{{c}}$, 
$b_{{c}}^{\dagger }b_{{c}}^{\dagger }b_{{c}}b_{{c}}$ 
and other operators of the class $[2,2]$ from the term ${\frac 12}
\left[ R,V\right] $, the ''good'' five-operator contributions of the $g^3$
-order $V_{good}^{(3)}\left( \alpha _c\right) \sim a_{{c}}^{\dagger }b_{{c}
}^{\dagger }b_{{c}}^{\dagger }b_{{c}}b_{{c}}$, $a_{{c}}^{\dagger } b_{{c}
}^{\dagger }b_{{c}}^{\dagger }b_{{c}}b_{{c}}$ and other operators of the
class $[3,2]$ from the term ${\frac 13}\left[ R,\left[ R,V\right] \right] $,
etc. Some of them will be interpreted below. 

In general, it is important to keep in mind that the so-called ''good''
terms and their H.c. counterparts, by definition, destroy the
no-clothed-particle and one-clothed-particle states (cf. Eq. (\ref
{KIProperty})). Unlike this, as mentioned in Subsect. 2.3, the bad terms
have not such property, viz., even if one of them destroys such clothed
states, its H.c. counterpart does not. Of course, among the operators
involved in the total Hamiltonian $H$ there are the operators of the class $%
[1,1]$ (for instance, $a_{{c}}^{\dagger }a_{{c}}$) that ensure these states
to be $H$ eigenstates.

As the next step of the clothing procedure we consider a recursive
continuation of Eq. (\ref{K_AfterFirstClothing}) with
\begin{equation}
K\left( \alpha _c\right) =H_F\left( \alpha _c\right) +V_C\left( \alpha
_c\right) =W^{\prime }K\left( \alpha _c^{\prime }\right) {W^{\prime }}%
^{\dagger }\equiv K^{\prime }\left( \alpha _c^{\prime }\right) =H_F\left(
\alpha _c^{\prime }\right) +V_C^{\prime }\left( \alpha _c^{\prime }\right) ,
\label{K_AfterSecondClothing}
\end{equation}
where the operators $\alpha _c^{\prime }$ in (\ref{K_AfterSecondClothing})
are related to the operators $\alpha _c$ in (\ref{K_AfterFirstClothing})
via the similarity transformation $\alpha _c^{\prime }={W^{\prime }}%
^{\dagger }\alpha _cW^{\prime }$. The interaction part $V_C\left( \alpha
_c\right) =H-H_F\left( \alpha _c\right) =W\left( \alpha _c\right) H\left(
\alpha _c\right) W^{\dagger }\left( \alpha _c\right)-H_F\left( \alpha _c\right)$ resulting from
the first clothing can be represented as
\[
V_C\left( \alpha _c\right) =\left. \exp (\frac d{d\lambda })H(R;\lambda
)\right| _{\lambda =0}-\,\,H_F\left( \alpha _c\right) ,
\]
where we follow the notation $A(B;\lambda )=e^{\lambda B}Ae^{-\lambda B}$
with the real parameter $\lambda $ for any operators $A$ and $B$.
Obviously, $A=A(B;0)$. According to this notation, the condition imposed
on the generator $R$ by Eq.~(\ref{EquationForR^1}) reads
\[
V+\left. \frac d{d\lambda }H_F(R;\lambda )\right| _{\lambda =0}=0 \, ,
\]
and the interaction generated by the first clothing UT acquires 
the expression
\[
V_C=\left. \exp (\frac d{d\lambda })\left[ M_{ren}(R;\lambda
)+V_{ren}(R;\lambda )\right] \right| _{\lambda =0}+
\]
\begin{equation}
\int\limits_0^1\,du\left. \left[ \exp (\frac d{d\lambda })-\exp (u\frac
d{d\lambda })\right] V(R;\lambda )\right| _{\lambda =0} \, .
\label{DiffFormFirstInteraction}
\end{equation}
This compact expression can be expanded in series
\[
V_C=M_{ren}+V_{ren}+\sum_{n=1}^\infty \left. \frac 1{n!}\frac{d^n}{d\lambda
^n}\left[ M_{ren}(R;\lambda )+V_{ren}(R;\lambda )+I_nV(R;\lambda )\right]
\right| _{\lambda =0}
\]
with the coefficients $I_n=\frac n{n+1}$. Keeping in mind the relationship
\begin{equation}
\left[ A\right] ^n\equiv \left. \frac{d^n}{d\lambda ^n}A(B;\lambda )\right|
_{\lambda =0}=\left[ B,\left[ B,...\left[ B,A\right] ...\right] \right]
\label{[B,A]MultipleCommutator}
\end{equation}
with $n$ brackets $(n=1,2,...)$ it is obvious that (\ref
{DiffFormFirstInteraction}) is a compact representation of $V_C$ as the
expansion in the multiple commutators $\left[ M_{ren}\right] ^n$, $\left[
V_{ren}\right] ^n$ and $\left[ V\right] ^n$. This expansion does not
contain any terms of the first order in the coupling constant.

The second clothing UT $W^{\prime }=\mathrm{exp}\left[ R^{(2)}\right] $ has
the generator $R^{(2)}$ which meets the condition
\[
V_{bad}^{(2)}+\left[ R^{(2)},H_F\right] =0
\]
or
\begin{equation}
V_{bad}^{(2)}+\left. \frac d{d\lambda }H_F(R^{(2)};\lambda )\right|
_{\lambda =0}=0  \label{EquationForR^2}
\end{equation}
to remove from $V_C$ all the $g^2$-order bad terms. The latter belong to the
classes $[2,0]$, $[2,1]$, $[3,0]$, $[3,1]$ and $[4,0]$. Their sum $%
V_{bad}^{(2)}$ enters in the decomposition $V_C^{(2)}={\frac 12}\left[
R,V\right] +M_{ren}^{(2)}\equiv V_{good}^{(2)}+V_{bad}^{(2)}$. In its turn, $%
V_C^{(2)}$ denotes the lowest order contributions to the series in $g$,
\[
V_C=V_{good}+V_{bad}=\sum_{n=2}^\infty V_C^{(n)}
\]
with
\[
V_{good}=V_{good}^{(2)}+V_{good}^{(3)}+V_{good}^{(4)}+\text{terms\thinspace
of\thinspace the }g^5\text{-and\thinspace higher\thinspace orders}
\]
and an analogous splitting of the operator $V_C(bad)$, whose terms do not
allow the no-clothed-particle and one-clothed-particle states to be $H$
eigenstates.

Further, repeating the tricks that result in Eq.\,(\ref
{DiffFormFirstInteraction}) for the after-first-clothing interactions and
using the condition (\ref{EquationForR^2}), we arrive to
\[
V_C^{\prime }\left( \alpha _c^{\prime }\right) =  H - H_F\left( \alpha
_c^{\prime }\right) = V_{good}(R^{(2)};\lambda) + V_{bad,rest} +
\]
\begin{equation}
\sum_{n=1}^{\infty} \left. \frac{1}{n!} \frac{d^n}{d\lambda ^n} \left[
V_{good}(R^{(2)};\lambda) + V_{bad,rest}(R^{(2)};\lambda) + I_n
V_{bad}^{(2)}(R^{(2)};\lambda) \right]\right|_{\lambda=0},
\label{DiffFormSecondInteraction}
\end{equation}
where $V_{bad,rest} = V_{bad} - V_{bad}^{(2)} = O\,(g^3) $. Along with the
framework interactions $V_{good}^{(n)}\, (n=2,3, \dots ) $ expressed through
the new clothed operators $\alpha _c^{\prime }$ this equation enables us to
evaluate the $g^n$-order corrections $(n=4,5, \dots )$ to them. As an
illustration, omitting the contributions of the $g^5$ - and higher orders we
find
\[
V_C^{\prime }\left( \alpha _c^{\prime }\right) = V_{good}^{(2)}\left( \alpha
_c^{\prime }\right) + V_{good}^{(3)}\left( \alpha _c^{\prime }\right) +
V_{bad}^{(3)}\left( \alpha _c^{\prime }\right) +
\]
\begin{equation}
\left[ R^{(2)},V_{good}^{(2)}\right]_{good} + {\frac 12}\left[
R^{(2)},V_{bad}^{(2)}\right]_{good} + V_{good}^{(4)}\left( \alpha _c^{\prime
}\right) + ...  \label{V_AfterSecondClothing}
\end{equation}
The r.h.s. of this equation involves the good interaction operators of
primary interest since they can be associated with different processes in
the meson-nucleon systems. In particular, as aforementioned, the term $%
V_{good}^{(2)}$ consists of the interactions of the class $[2,2]$. In order
to find the corrections of the $g^4$-order to them one needs to separate out
the operators of the same class of the three last terms in the r.h.s. of (%
\ref{V_AfterSecondClothing}). We have pointed out the $g^3$-order
contribution $V_{bad}^{(3)}$ as the first candidate for eliminating via the
third clothing UT.

\subsection{Details of Calculations}

We have seen how the interactions between clothed (physical) particles can
be constructed when handling the multiple commutators $\left[ V\right]^n$
\,(n=1,2,...) with $n$-brackets. In the framework of the Yukawa model or any
other field model with a polynomial interaction the operator $V$ can be
represented in the following symbolic form:
\begin{equation}
V \equiv f*m+H.c.,  \label{V_badInGeneralForm}
\end{equation}
where $f*m$ is a polynomial composed of products of fermionic and mesonic
operators.

In general, to obtain recursive relations for the commutators of increasing
complexity, it is convenient to write in accordance with Eq.(\ref
{[B,A]MultipleCommutator})
\begin{equation}
\left[ V\right] ^n=\lim_{\lambda \rightarrow 0}{\frac{d^n}{d\lambda ^n}}%
\left( e^{\lambda R}f*me^{-\lambda R}\right) +H.c.\,.
\label{[R,V]MultipleCommutatorDifferentialForm}
\end{equation}

Taking into account that $R$ is an antihermitian operator, we have:
\begin{equation}
\left[ V\right] ^n=\lim_{\lambda \rightarrow 0}{\frac{d^n}{d\lambda ^n}}%
\left( f(\lambda )*m(\lambda )\right) +H.c..
\label{[R,V]MultipleCommutatorDifferentialForm1}
\end{equation}
For brevity, we assume $f(R;\lambda )\equiv f(\lambda )$ and $m(R;\lambda
)\equiv m(\lambda )$.

Then, using the Leibnitz formula
\[
{\frac{d^n}{d\lambda ^n}}\left( f(\lambda )*m(\lambda )\right)
=\sum_{s=0}^nC_n^s[f(\lambda )]^{n-s}*[m(\lambda )]^s,
\]
we find
\[
\lim_{\lambda \rightarrow 0}{\frac{d^n}{d\lambda ^n}}\left( f(\lambda
)*m(\lambda )\right) =\sum_{s=0}^nC_n^s[f(0)]^{n-s}*[m(0)]^s,
\]
whence

\begin{equation}
\left[ V_{bad}\right] ^n=\sum_{s=0}^nC_n^s[f]^{n-s}*[m]^s+H.c..
\label{[R,V]MultipleCommutatorDifferentialFinalForm}
\end{equation}

At the first stage of our procedure for the Yukawa model

\begin{equation}
V_{bad}=V\left( \alpha _c\right) =\int d\mathbf{k}f(\mathbf{k})m(\mathbf{k}%
)+H.c.,\,\,\,\,R=R_1\left( \alpha _c\right) =\int d\mathbf{k}\hat{R}_c^{%
\mathbf{k}}\,m(\mathbf{k})-H.c.,
\end{equation}
with $f(\mathbf{k})=\hat{V}_c^{\mathbf{k}}$ and $m(\mathbf{k})=a_c(\mathbf{k}%
)$. For such $V_{bad}$ formula (\ref
{[R,V]MultipleCommutatorDifferentialFinalForm}), where $[f]^{n-s}*[m]^s%
\equiv \int d\mathbf{k}[f(\mathbf{k})]^{n-s}[m(\mathbf{k})]^s$, yields

\begin{equation}
\left[ V_{bad}\right] ^1=\int d\mathbf{k}\left[ [f(\mathbf{k})]^1m(\mathbf{k}%
)+f(\mathbf{k})[m(\mathbf{k})]^1\right] +H.c.,
\label{[R,V]MultipleCommutatorFor_n=1}
\end{equation}

\begin{equation}
\left[ V_{bad} \right]^2 = \int d \mathbf{k} \left[ [f(\mathbf{k})]^{2} m(%
\mathbf{k}) + 2 [f(\mathbf{k})]^{1} [m(\mathbf{k})]^{1} + f(\mathbf{k}) [m(%
\mathbf{k})]^{2} \right] + H.c.,  \label{[R,V]MultipleCommutatorFor_n=2}
\end{equation}

From these equations it follows that

\begin{eqnarray}
\left[ R,V \right] = \left[ V \right]^1 = \int d \mathbf{k}_1 d \mathbf{k}_2
\left\{ \left[ \hat {R}^{\mathbf{k}_2}, \hat {V}^{\mathbf{k}_1} \right] a(%
\mathbf{k}_2) a(\mathbf{k}_1) \right.  \nonumber \\
\left. - \left[ \hat {R}^{\mathbf{k}_2 \dagger}, \hat {V}^{\mathbf{k}_1}
\right] a^\dagger(\mathbf{k}_2) a(\mathbf{k}_1) + \hat {V}^{\mathbf{k}_1}
\hat {R}^{\mathbf{k}_2 \dagger} \delta( \mathbf{k}_1 - \mathbf{k}_2 )
\right\} + H.c.  \label{[R,V]ExplicitForm}
\end{eqnarray}

\begin{eqnarray}
\left[ R, \left[ R,V \right] \right]= \left[ V \right]^2 = \int d \mathbf{k}%
_1 d \mathbf{k}_2 d \mathbf{k}_3 \left\{ A_1( \mathbf{k}_1, \mathbf{k}_2,
\mathbf{k}_3 ) a^\dagger( \mathbf{k}_2 ) a^\dagger( \mathbf{k}_1 ) a(
\mathbf{k}_3 ) \right.  \nonumber \\
+ A_2( \mathbf{k}_1, \mathbf{k}_2, \mathbf{k}_3 ) a^\dagger( \mathbf{k}_2 )
a( \mathbf{k}_1 ) a( \mathbf{k}_3 ) + A_3( \mathbf{k}_1, \mathbf{k}_2,
\mathbf{k}_3 ) a( \mathbf{k}_2 ) a( \mathbf{k}_1 ) a( \mathbf{k}_3 )
\nonumber \\
\left. + A_4( \mathbf{k}_1, \mathbf{k}_2 ) a^\dagger( \mathbf{k}_2 ) \delta(
\mathbf{k}_1 - \mathbf{k}_3 ) + A_5( \mathbf{k}_1, \mathbf{k}_2 ) a( \mathbf{%
k}_2 ) \delta( \mathbf{k}_1 - \mathbf{k}_3 ) \right\} + H.c.,  \nonumber
\end{eqnarray}

\begin{eqnarray}
A_1( \mathbf{k}_1, \mathbf{k}_2, \mathbf{k}_3 ) = \left[ \hat {R}^{\mathbf{k}%
_3}, \left[ \hat {R}^{\mathbf{k}_1}, \hat {V}^{\mathbf{k}_2} \right]
^\dagger \right], A_2( \mathbf{k}_1, \mathbf{k}_2, \mathbf{k}_3 ) = \left[
\hat {R}^{\mathbf{k}_3}, \left[ \hat {R}^{\mathbf{k}_1}, \hat {V}^{\mathbf{k}%
_2 \dagger} \right] + \left[ \hat {R}^{\mathbf{k}_2}, \hat {V}^{\mathbf{k}_1
\dagger} \right] ^\dagger \right],  \nonumber \\
A_3( \mathbf{k}_1, \mathbf{k}_2, \mathbf{k}_3 ) = \left[ \hat {R}^{\mathbf{k}%
_3}, \left[ \hat {R}^{\mathbf{k}_1}, \hat {V}^{\mathbf{k}_2} \right]
\right], A_4( \mathbf{k}_1, \mathbf{k}_2 ) = \hat {R}^{\mathbf{k}_1} \left\{
\left[ \hat {R}^{\mathbf{k}_1}, \hat {V}^{\mathbf{k}_2} \right] ^\dagger +
\left[ \hat {R}^{\mathbf{k}_2}, \hat {V}^{\mathbf{k}_1} \right] ^\dagger
\right\},  \nonumber
\end{eqnarray}

\begin{eqnarray}
A_5(\mathbf{k}_1,\mathbf{k}_2) &=&\hat{R}^{\mathbf{k}_1}\left[ \hat{R}^{%
\mathbf{k}_1},\hat{V}^{\mathbf{k}_2\dagger }\right] ^{\dagger }+\left[ \hat{R%
}^{\mathbf{k}_2},\hat{R}^{\mathbf{k}_1}\right] \hat{V}^{\mathbf{k}_1\dagger }
\nonumber \\
&&+2\hat{R}^{\mathbf{k}_1}\left[ \hat{R}^{\mathbf{k}_2},\hat{V}^{\mathbf{k}%
_1\dagger }\right] +\left[ \hat{R}^{\mathbf{k}_2},\hat{V}^{\mathbf{k}%
_1}\right] \hat{R}^{\mathbf{k}_1\dagger }+\hat{V}^{\mathbf{k}_1}\left[ \hat{R%
}^{\mathbf{k}_2},\hat{R}^{\mathbf{k}_1\dagger }\right] .
\label{[R,[R,V]]ExplicitForm}
\end{eqnarray}
For brevity, the superscript $c$ has been omitted in the r.h.s. of Eqs. (\ref
{[R,V]ExplicitForm}) and (\ref{[R,[R,V]]ExplicitForm}).

Performing the normal ordering of the fermion operators in Eqs. (\ref
{[R,V]ExplicitForm}) and (\ref{[R,[R,V]]ExplicitForm}), we get a simple
recipe to select the $2\longleftrightarrow 2$ and $2\longleftrightarrow 3$
interaction operators of the $g^2$- and $g^3$-orders between the partially
clothed pions, nucleon and antinucleons (in particular, $\pi N\rightarrow
\pi N$, $NN\rightarrow NN$ and $NN\leftrightarrow \pi NN$). At the same time
this algebraic technique enables to select the two-operator (one-body)
contributions which cancel the meson and fermion mass counterterms $%
M_{ren}(mes)$ and $M_{ren}(ferm)$ in the $g^2$-order (details are in \cite
{SheShi2001} and \cite{KordaShebekoPhys.Rev.2004}). In addition, we
encounter the three-operator (vertex-like) ''radiative'' corrections which
together with the similar terms from the commutators $\left[
R,M_{ren}(mes)\right] $ and $\left[ R,M_{ren}(ferm)\right] $ cancel the
''charge'' counterterm $V_{ren}$ in the $g^3$-order. The remaining bad terms
must be removed via successive clothing UT's.

Along this guideline, we arrive at the decomposition:
\begin{eqnarray}
K(\alpha _c)&=&K_F(\alpha _c)+K(NN\rightarrow NN)+K(\bar{N}\bar{N}
\rightarrow \bar{N}\bar{N})+K(N\bar{N}\rightarrow N\bar{N})  \nonumber \\
&+&K(\pi N \rightarrow \pi N)+K(\pi \bar{N}\rightarrow \pi \bar{N})+K(\pi
\pi \leftrightarrow N\bar{N})  \nonumber \\
&+& K(NN \leftrightarrow \pi NN)+K(\bar{N}\bar{N}\leftrightarrow \pi \bar{N}
\bar{N})+K(N\bar{N}\leftrightarrow \pi N\bar{N})  \nonumber \\
&+&K(N\bar{N} \leftrightarrow \pi \pi \pi )+K(\pi N\leftrightarrow \pi \pi
N)+K(\pi \bar{N}\leftrightarrow \pi \pi \bar{N}) + \dots
\label{ClothedHamiltonianUpToG^3Order}
\end{eqnarray}
where the interactions between the clothed nucleons ($N$), antinucleons ($%
\bar{N}$) and pions ($\pi $) have been separated out.

\subsection{Pion-nucleon interaction operator}

To obtain the explicit expression for the operator $K(\pi N\rightarrow \pi
N) $ one needs to separate out the $b_c^{\dagger }b_c$-type terms form the
commutator $\left[ \hat{R}^{\mathbf{k}_2\dagger },\hat{V}^{\mathbf{k}%
_1}\right] $ and its H.c. part in the r.h.s. of Eq. (\ref{[R,V]ExplicitForm}%
). The result can be represented in the following covariant (Feynman-like)
form:

\[
K(\pi N\rightarrow \pi N)=\int d\mathbf{p}_1d\mathbf{p}_2d\mathbf{k}_1d%
\mathbf{k}_2\,V_{\pi N}(\mathbf{k}_2,\mathbf{p}_2;\mathbf{k}_1,\mathbf{p}%
_1)a_c^{\dagger }(\mathbf{k}_2)b_c^{\dagger }(\mathbf{p}_2)a_c(\mathbf{k}%
_1)b_c(\mathbf{p}_1),
\]
\[
V_{\pi N}(\mathbf{k}_2,\mathbf{p}_2;\mathbf{k}_1,\mathbf{p}_1)=\frac{g^2}{%
2(2\pi )^3}\frac m{\sqrt{\omega _{\mathbf{k}_1}\omega _{\mathbf{k}_2}E_{%
\mathbf{p}_1}E_{\mathbf{p}_2}}}\delta (\mathbf{p}_1+\mathbf{k}_1-\mathbf{p}%
_2-\mathbf{k}_2)
\]
\begin{equation}
\times \bar{u}(\mathbf{p}_2)\left\{ \frac 12\left[ \frac 1{\not{p}_1+\not%
{k}_1+m}+\frac 1{\not{p}_2+\not{k}_2+m}\right] +\frac 12\left[ \frac 1{\not%
{p}_1-\not{k}_2+m}+\frac 1{\not{p}_2-\not{k}_1+m}\right] \right\} u(\mathbf{p%
}_1).  \label{PiNInteraction}
\end{equation}
Henceforth the spin indices are omitted and all summations over them are
implied.

The $\pi N$ quasipotential being the kernel of an approximate integral
equation (cf. Eq. (4.9) in \cite{SheShi2001}) for the $\pi N$ scattering
wave function in momentum space is determined as in \cite{SheShi2001}:

\begin{equation}
\tilde{V}_{\pi N}(\mathbf{k}_2,\mathbf{p}_2;\mathbf{k}_1,\mathbf{p}%
_1)=\left\langle a_c^{\dagger }(\mathbf{k}_2)b_c^{\dagger }(\mathbf{p}%
_2)\Omega \mathbf{|}K(\pi N\rightarrow \pi N)\mathbf{|}a_c^{\dagger }(%
\mathbf{k}_1)b_c^{\dagger }(\mathbf{p}_1)\Omega \right\rangle
\label{PiNQuasipotential}
\end{equation}
and it turns out to be equal to $V_{\pi N}(\mathbf{k}_2,\mathbf{p}_2;\mathbf{%
k}_1,\mathbf{p}_1)$.

In order to interpret Eq.~(\ref{PiNInteraction}), we write an intermediate
analytical result that leads to it, namely
\[
V_{\pi N}(\mathbf{k}_2,\mathbf{p}_2;\mathbf{k}_1,\mathbf{p}_1)=\frac{g^2}{%
2(2\pi )^3}\frac m{\sqrt{\omega _{\mathbf{k}_1}\omega _{\mathbf{k}_2}E_{%
\mathbf{p}_1}E_{\mathbf{p}_2}}}\delta (\mathbf{p}_1+\mathbf{k}_1-\mathbf{p}%
_2-\mathbf{k}_2)
\]
\begin{equation}
\times \frac 12\bar{u}(\mathbf{p}_2)\gamma _5\left\{ P\left( \mathbf{k}_2,%
\mathbf{p}_2;\mathbf{k}_1,\mathbf{p}_1\right) +P\left( \mathbf{k}_1,\mathbf{p%
}_1;\mathbf{k}_2,\mathbf{p}_2\right) \right\} \gamma _5u(\mathbf{p}_1),
\label{PiNInteraction_1}
\end{equation}
\[
P\left( \mathbf{k}_2,\mathbf{p}_2;\mathbf{k}_1,\mathbf{p}_1\right) =\frac
m{E_{\mathbf{q}}}\left[ \frac{\Lambda _{+}\left( q\right) }{E_{\mathbf{q}%
}-E_{\mathbf{p}_1}-\omega _{\mathbf{k}_1}}+\frac{\Lambda _{-}^{\dagger
}\left( q\right) }{E_{\mathbf{q}}+E_{\mathbf{p}_1}+\omega _{\mathbf{k}_1}}%
\right]
\]
\begin{equation}
-\frac m{E_{\mathbf{q}^{\prime }}}\left[ \frac{\Lambda _{+}\left( q^{\prime
}\right) }{E_{\mathbf{p}_2}-E_{\mathbf{q}^{\prime }}-\omega _{\mathbf{k}_1}}-%
\frac{\Lambda _{-}^{\dagger }\left( q^{\prime }\right) }{E_{\mathbf{p}_2}+E_{%
\mathbf{q}^{\prime }}-\omega _{\mathbf{k}_1}}\right]   \label{PiNPropagator}
\end{equation}
with the standard notations $\Lambda _{+}\left( \Lambda _{-}\right) $ for
the projection operators on the fermion positive (negative)-energy states $%
\Lambda _{\pm }\left( q\right) =\left( \pm \not{q}+m\right) /2m$.

Separate contributions to the r.h.s. of Eq. (\ref{PiNPropagator}) can be
represented as in Fig.1, viz., via graphs $a$ and $b$, which correspond to
the two terms in the first square brackets, and $c$ and $d$ associated with
the two ones in the second square brackets. The translational invariance
imposes the constraints $\mathbf{k}_1+\mathbf{p}_1=\mathbf{q}=\mathbf{k}_2+%
\mathbf{p}_2$ and $\mathbf{p}_1-\mathbf{k}_2=\mathbf{q}^{\prime }=\mathbf{p}%
_2-\mathbf{k}_1$ upon the momenta involved. In other words, it implies that
three-momentum is conserved in the each vertex of these graphs.

\begin{figure}[ht]
\scalebox{0.5}{\includegraphics{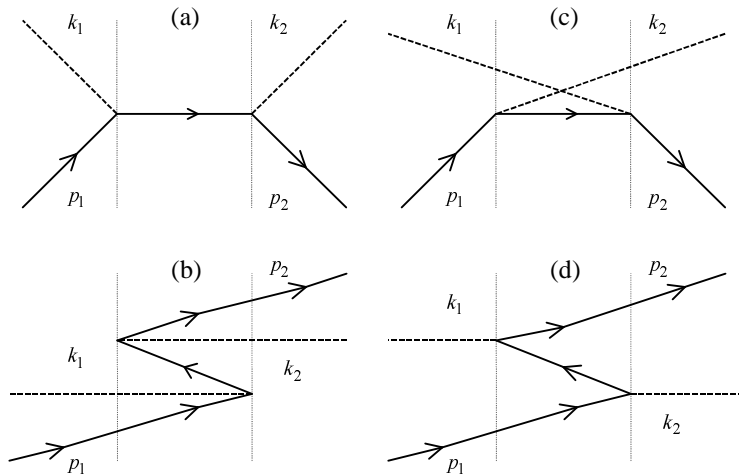}}
\caption{\label{labalfig1}
Different contributions to the $\pi N$ quasipotential within the
Old-Fashioned Perturbation Theory (OFPT).
}
\end{figure}


Such graphs are well-known from the old-fashioned perturbation theory
(OFPT), where, for example, the $\pi N$ scattering $T$-matrix in the $g^2$%
-order is determined by the elements
\begin{eqnarray*}
\left\langle \mathbf{k}_2,\mathbf{p}_2\mathbf{|}T^{\left( 2\right) }(E+i0)%
\mathbf{|k}_1,\mathbf{p}_1\right\rangle &=&\left\langle \mathbf{k}_2,\mathbf{%
p}_2\mathbf{|}V(E+i0-H_0)^{-1}V\mathbf{|k}_1,\mathbf{p}_1\right\rangle \\
&=&\sum\limits_i\left\langle \mathbf{k}_2,\mathbf{p}_2\mathbf{|}V\mathbf{|}%
i\right\rangle (E+i0-E_i)^{-1}\left\langle i\mathbf{|}V\mathbf{|k}_1,\mathbf{%
p}_1\right\rangle .
\end{eqnarray*}
Here $E$ the collision energy and the summation is performed over all
permissible intermediate states (the $H_0$ eigenstates) with energies $E_i$.
In this context, the inverse energy denominators in the r.h.s. of Eq. (\ref
{PiNPropagator}) have the form $(E-E_i)^{-1}$ with the appropriate values of
$E$ and $E_i$. For example, $\left( E_{\mathbf{q}}+E_{\mathbf{p}_1}+\omega _{%
\mathbf{k}_1}\right) ^{-1}$ and $\left( E_{\mathbf{p}_2}+E_{\mathbf{q}%
^{\prime }}-\omega _{\mathbf{k}_1}\right) ^{-1}$ may be related to the
noncovariant propagators
\begin{eqnarray*}
D^{-1}\left( E\right) |_{E=E_{\mathbf{p}_2}+\omega _{\mathbf{k}_2}} &\equiv
&\left( E-E_{\mathbf{p}_1}-\omega _{\mathbf{k}_1}-E_{\mathbf{p}_2}-\omega _{%
\mathbf{k}_2}-E_{\mathbf{q}}\right) ^{-1}|_{E=E_{\mathbf{p}_2}+\omega _{%
\mathbf{k}_2}} \\
&=&-\left( E_{\mathbf{q}}+E_{\mathbf{p}_1}+\omega _{\mathbf{k}_1}\right)
^{-1},
\end{eqnarray*}
and
\begin{eqnarray*}
\overline{D}^{-1}\left( E\right) |_{E=E_{\mathbf{p}_1}+\omega _{\mathbf{k}%
_1}} &\equiv &\left( E-E_{\mathbf{p}_1}-E_{\mathbf{p}_2}-E_{\mathbf{q}%
^{\prime }}\right) ^{-1}|_{E=E_{\mathbf{p}_1}+\omega _{\mathbf{k}_1}} \\
&=&-\left( E_{\mathbf{p}_2}+E_{\mathbf{q}^{\prime }}-\omega _{\mathbf{k}%
_1}\right) ^{-1},
\end{eqnarray*}
being associated with the five and three internal lines putting in between
the dotted verticals, respectively, in graphs $b$ and $d$. Both of them
contain the $N\overline{N}$ pair in the intermediate states. Note that we
have to distinguish the intermediate states of clothed particles in our
approach from the intermediate states of bare particles with physical masses
that usually appear in OFPT.

In addition, we would like to note that the graphs in Fig. 1 are
topologically equivalent to the four time-ordered Feynman diagrams displayed
in Fig. 20 in Chapter 13 of Schweber's book \cite{SchweberBook}. However, in
the Schr\"{o}dinger picture used here, where all events are related to one
and the same instant $t=0$ the use of such an analogy seems to be
misleading. In fact, the line directions in Fig. 1 are given with the sole
scope to discriminate between the nucleon and antinucleon states.

The Feynman-like propagators in Eq. (\ref{PiNInteraction}) appear after
trivial transformations, e.g., adding the contribution $a$ and $b$, it is
readily seen that
\[
\frac m{E_{\mathbf{q}}}\left[ \frac{\Lambda _{+}\left( q\right) }{E_{\mathbf{%
q}}-E_{\mathbf{p}_1}-\omega _{\mathbf{k}_1}}+\frac{\Lambda _{-}^{\dagger
}\left( q\right) }{E_{\mathbf{q}}+E_{\mathbf{p}_1}+\omega _{\mathbf{k}_1}}%
\right] =-\frac 1{\not{s}_{_1}-m},
\]
where we have denoted the ''left'' Mandelstam vector $s_1=\left( E_{\mathbf{p%
}_1}+\omega _{\mathbf{k}_1},\mathbf{p}_1+\mathbf{k}_1\right) =p_1+k_1$. Now,
taking into account the property $\gamma _5\gamma _\mu \gamma _5=-\gamma
_\mu $, one shortly derives Eq. (\ref{PiNInteraction}).

We emphasize that energy conservation is not assumed in constructing this
and other quasipotentials, and this forces us to give much more detailed and
complicated expressions for these quasi-potentials. Indeed, the coefficients
for the $a_c^{\dagger }b_c^{\dagger }a_cb_c$-terms appearing in the $K(\pi
N\rightarrow \pi N)$ expansion generally do not fulfill the on-energy-shell
condition
\begin{equation}
E_{\mathbf{p}_1}+\omega _{\mathbf{k}_1}=E_{\mathbf{p}_2}+\omega _{\mathbf{k}%
_2},  \label{PiNEnergyShell}
\end{equation}
In this connection, the ``left'' four-vector $s_1$ is not necessarily equal
to the ''right'' Mandelstam vector $s_2=p_2+k_2$. Clearly, if the condition (%
\ref{PiNEnergyShell}) is fulfilled, then
$p_1-k_2=u$, so that the first (second) half-sum of the covariant
propagators in the r.h.s. of Eq. (\ref{PiNInteraction}) is converted into
the $s$-pole ($u$-pole) propagator typical of Feynman technique. Thus, only
the corresponding on-energy-shell part of the $\pi N$ quasipotential is
represented in Fig. 2 by the genuine Feynman diagram.

\begin{figure}[ht]
\scalebox{0.5}{\includegraphics{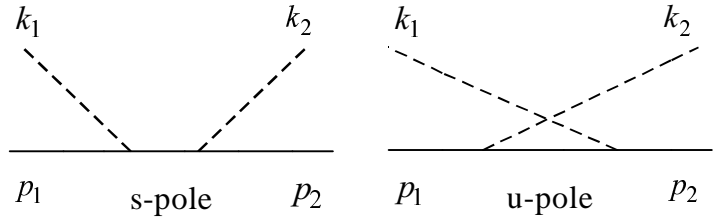}}
\caption{\label{labalfig2}
On-energy-shell contributions to the $\pi N$-quasipotential.}
\end{figure}

The quasipotential is nonlocal since the vertex factors and propagators in
Eq. (\ref{PiNInteraction}) are dependent not only on the relative
three-momenta involved but also on their total three-momentum. Besides, this
embodies the nonstatic (recoil) effects in all orders of the so-called $%
1/c^2 $- expansion (see \cite{FoldyKrajcik}).

In the static limit, one can show that the off-energy-shell contributions
differ from the on-energy-shell results by the values of the order $\mathbf{q%
}\,_r^2/\eta ^2$ where $\mathbf{q}_r^{}$ is the relative $\pi N$ momentum
and $\eta =m\mu /(m+\mu )$ is the reduced $\pi N$ mass. One needs additional
investigations to show to what extent these differences can be neglected when
describing $\pi N$ scattering .

To our knowledge the formula (\ref{PiNQuasipotential}) (albeit given in
other form) has been first derived in \cite{ShirokovVisinescuRevRoumPhys1974}
within the clothed particle approach.


The quasipotential determined by Eq. (\ref{PiNInteraction}) has the same
Feynman-like structure as the effective $\pi N$ interaction obtained in \cite
{SatoLee.Phys.Rev.C54.2660.1996} (see the final formula below Eq. (A.8)
therein). As mentioned in Introduction, the authors of \cite
{SatoLee.Phys.Rev.C54.2660.1996} have used the other realization of UT
method, which is similar to the Fr\"{o}hlich transformation (see, e.g., \cite
{Davydov1973}) in the theory of metals.

In addition, one can show that the coefficients $V_{\pi N}(\mathbf{k}_2,%
\mathbf{p}_2;\mathbf{k}_1,\mathbf{p}_1)$ in Eq. (\ref{PiNInteraction}) do
not differ from those obtained in Ref. \cite{Fuda.Ann.Phys.231.1.1994}
(after some simplifications of the model Hamiltonian used there). However,
the corresponding pion-nucleon interaction operator from Ref.~\cite
{Fuda.Ann.Phys.231.1.1994} contains the bare particle $a^{\dagger
}b^{\dagger }ab$-terms instead of the clothed particle $a_c^{\dagger
}b_c^{\dagger }a_cb_c$-ones in our expression for $K(\pi N\rightarrow \pi N)$%
. It would be interesting and worthwhile to check if the same coincidence of
results will be obtained for higher-order terms in the coupling constant.

\subsection{Nucleon-nucleon interaction operator}

After normal ordering of the fermion operators in the third integral in the
r.h.s. of Eq. (\ref{[R,V]ExplicitForm}) and its Hermitian conjugate part the
$N N \rightarrow N N $ interaction operator can be represented in the form:

\[
K(NN\rightarrow NN)=\int d\mathbf{p }_1 d\mathbf{p }_2 d\mathbf{p }%
_1^{\prime } d\,\mathbf{p }_2^{\prime } V_{NN}(\mathbf{p }_1^{\prime },
\mathbf{p}_2^{\prime }; \mathbf{p}_1, \mathbf{p}_2) b_c^{\dagger } (\mathbf{%
p }_1^{\prime }) b_c^{\dagger }(\mathbf{p }_2^{\prime }) b_c(\mathbf{p }_1)
b_c(\mathbf{p }_2),
\]
\[
\smallskip V_{N N}(\mathbf{p }_1^{\prime }, \mathbf{p}_2^{\prime }; \mathbf{p%
}_1, \mathbf{p}_2) = - \frac{1}{2} \frac{g^{2}}{(2\pi )^{3}}\frac{1}{%
2\omega_{\mathbf{p }_1^{\prime }-\mathbf{p }_1}} \frac{m^2}{\sqrt{E_{\mathbf{%
p}_1}E_{\mathbf{p}_2} E_{\mathbf{p}_{1}^{\prime }}E_{\mathbf{p}_{2}^{\prime
}} } } \delta (\mathbf{p }_1^{\prime }+\mathbf{p}_2^{\prime } -\mathbf{p}_1-%
\mathbf{p}_2)
\]

\begin{equation}
\times \bar{u}(\mathbf{p }_1^{\prime } )\gamma _{5}u(\mathbf{p }_1) \left\{
\frac{1}{ E_{\mathbf{p}_1 } - E_{\mathbf{p}_1^{\prime } } - \omega_{\mathbf{%
p }_1^{\prime }-\mathbf{p }_1 } } + \frac{1}{E_{\mathbf{p}_1^{\prime } } -
E_{\mathbf{p}_1 } - \omega_{\mathbf{p }_1^{\prime }-\mathbf{p }_1 } }
\right\} \bar{u}(\mathbf{p }_2^{\prime } )\gamma_{5} u(\mathbf{p }_2)
\label{OFPTNNInteraction}
\end{equation}

As in case of the $\pi N$, we encounter here the noncovariant
(``nonrelativistic'') propagators linear in the intermediate energies. It
allows us to relate the two graphs in Fig.~3 to the r.h.s. of Eq. (\ref
{OFPTNNInteraction}).

\begin{figure}[ht]
\scalebox{0.5}{\includegraphics{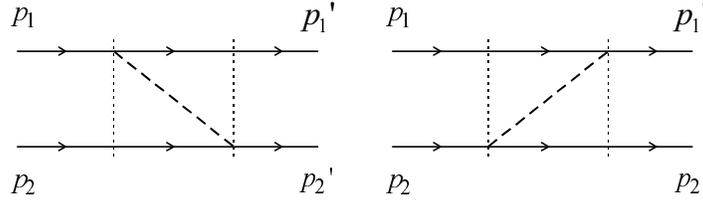}}
\caption{\label{labalfig3}
Illustration of Eq.(\ref{OFPTNNInteraction}) as a sum of two
OFPT diagrams with the intermediate pion (dashed lines) on its mass shell.}
\end{figure}


The left and right diagrams in Fig.3 are associated with the OFPT
denominators,
\begin{eqnarray*}
D\left( E\right) |_{E=E_{\mathbf{p}_1}+E_{\mathbf{p}_2}} &\equiv &\left(
E-E_{\mathbf{p}_1^{\prime }}-E_{\mathbf{p}_2}-\omega _{\mathbf{p}_1^{\prime
}-\mathbf{p}_1}\right) |_{E=E_{\mathbf{p}_1}+E_{\mathbf{p}_2}} \\
&=&E_{\mathbf{p}_1}-E_{\mathbf{p}_1^{\prime }}-\omega _{\mathbf{p}_1^{\prime
}-\mathbf{p}_1}
\end{eqnarray*}
and
\begin{eqnarray*}
\overline{D}\left( E\right) |_{E=E_{\mathbf{p}_1^{\prime }}+E_{\mathbf{p}%
_2^{\prime }}} &\equiv &\left( E-E_{\mathbf{p}_1}-E_{\mathbf{p}_2^{\prime
}}-\omega _{\mathbf{p}_2^{\prime }-\mathbf{p}_2}\right) |_{E=E_{\mathbf{p}%
_1^{\prime }}+E_{\mathbf{p}_2^{\prime }}} \\
&=&E_{\mathbf{p}_1^{\prime }}-E_{\mathbf{p}_1}-\omega _{\mathbf{p}_1-\mathbf{%
p}_1^{\prime }},
\end{eqnarray*}
respectively (cf. discussion in Sect. 3.2).

Though Eq.(\ref{OFPTNNInteraction}) is equivalent to
\[
V_{N N}(\mathbf{p}_1^{\prime }, \mathbf{p}_2^{\prime }; \mathbf{p}_1,
\mathbf{p}_2) = -\frac{1}{2} \frac{g^{2}}{(2\pi )^3} \frac{m^2}{\sqrt{E_{%
\mathbf{p}_1}E_{\mathbf{p}_{2}} E_{\mathbf{p}_{1}^{\prime }}E_{\mathbf{p}%
_{2}^{\prime }} } } \delta (\mathbf{p}_1^{\prime }+\mathbf{p}_2^{\prime } -%
\mathbf{p}_1-\mathbf{p}_2)
\]
\begin{equation}
\times\bar{u}(\mathbf{p }_1^{\prime }) \gamma _{5}u(\mathbf{p }_1) \frac{1}{%
(p_{1}-p_{1}^{\prime })^{2}-\mu ^{2} } \bar{u}(\mathbf{p }_2^{\prime }
)\gamma _{5} u(\mathbf{p }_2),  \label{NNInteraction}
\end{equation}
the occurrence of such primary denominators is typical of the
three-dimensional formalism developed here.

The corresponding relativistic and properly symmetrized $NN$ interaction, is
given by the quasipotential
\begin{eqnarray}
\tilde{V}_{NN}(\mathbf{p}_1^{\prime },\mathbf{p}_2^{\prime };\mathbf{p}_1,%
\mathbf{p}_2) &=&\left\langle b_c^{\dagger }(\mathbf{p}_1^{\prime
})b_c^{\dagger }(\mathbf{p}_2^{\prime })\Omega \mathbf{|}K(NN\rightarrow NN)%
\mathbf{|}b_c^{\dagger }(\mathbf{p}_1)b_c^{\dagger }(\mathbf{p}_2)\Omega
\right\rangle  \nonumber \\
&=&\frac 12\left\{ V_{NN}(\mathbf{p}_1^{\prime },\mathbf{p}_2^{\prime };%
\mathbf{p}_1,\mathbf{p}_2)-V_{NN}(\mathbf{p}_1^{\prime },\mathbf{p}%
_2^{\prime };\mathbf{p}_2,\mathbf{p}_1)\right.  \nonumber \\
&&\left. -V_{NN}(\mathbf{p}_2^{\prime },\mathbf{p}_1^{\prime };\mathbf{p}_1,%
\mathbf{p}_2)+V_{NN}(\mathbf{p}_2^{\prime },\mathbf{p}_1^{\prime };\mathbf{p}%
_2,\mathbf{p}_1)\right\} ,  \label{NNQuasipotential}
\end{eqnarray}
(see formula (4.16) from Ref. \cite{SheShi2001}).

Substituting (\ref{NNInteraction}) into (\ref{NNQuasipotential}), we express
the quasipotential of interest through the covariant (Feynman-like)
``propagators'',
\[
\tilde V_{N N}(\mathbf{p }_1^\prime, \mathbf{p }_2^\prime ; \mathbf{p }_1,%
\mathbf{p }_2) = - \frac{1}{2} \frac{ g^{2}} {(2\pi )^{3}}\frac{m^{2}}{2
\sqrt{E_{\mathbf{p}_1}E_{\mathbf{p}_{2}} E_{\mathbf{p} _{1}^{\prime }}E_{%
\mathbf{p}_{2}^{\prime }} } } \delta (\mathbf{p }_1^{\prime }+\mathbf{p}%
_2^{\prime } -\mathbf{p}_1-\mathbf{p}_2)
\]
\begin{equation}
\times \bar{u}(\mathbf{p }_1^{\prime })\gamma _{5}u(\mathbf{p }_1) \frac{1}{2%
} \left\{ \frac{1}{(p_{1}-p_{1}^{\prime })^{2}-\mu ^{2}} + \frac{1}{%
(p_{2}-p_{2}^{\prime })^{2}-\mu ^{2}} \right\} \bar{u}(\mathbf{p }_2^{\prime
} )\gamma_{5} u(\mathbf{p }_2) - (1 \leftrightarrow 2).
\label{ClothedNNInteraction}
\end{equation}

The r.h.s. of this equation consists of the ``direct'' term written
explicitly and the ``exchange'' one $(1 \leftrightarrow 2)$ with the
prescription that the variables with label 1 and 2 should be mutually
interchanged. 
Formula (\ref{ClothedNNInteraction}) determines the $NN$ part of an
one-boson-exchange interaction derived via the Okubo transformation method
in \cite{KorchinShebekoPhysAtNucl1993} (see also \cite
{FudaZhang.Phys.Rev.C51.23.1995} ) taking into account the pion and
heavy-meson exchanges.

As has been emphasized in Ref.\cite{KorchinShebekoPhysAtNucl1993}, a
distinctive feature of potential (\ref{ClothedNNInteraction}) is the
presence of a covariant (Feynman-like) ``propagator''
\begin{equation}
\frac 12\left\{ \frac 1{(p_1-p_1^{\prime })^2-\mu ^2}+\frac
1{(p_2-p_2^{\prime })^2-\mu ^2}\right\} \, .  \label{propagator}
\end{equation}
On the energy shell for the $NN$ scattering, that is
\begin{equation}
E_i\equiv E_{\mathbf{p}_1}+E_{\mathbf{p}_2}=E_{\mathbf{p}_1^{\prime }}+E_{%
\mathbf{p}_2^{\prime }}\equiv E_f,  \label{NNEnergyShell}
\end{equation}
this expression is converted into the genuine Feynman propagator which
occurs when evaluating the $S$ - matrix for $NN$ scattering in the $g^2$%
-order. The corresponding graphs are displayed in Fig. 4. The quasipotential
$\tilde{V}_{NN}$ can be related to these Feynman diagrams, being different
from the corresponding Feynman $NN$-scattering amplitude in two respects,
namely, $\tilde{V}_{NN}$ does not contain the factor $\delta (E_{\mathbf{p}%
_1}+E_{\mathbf{p}_2}-E_{\mathbf{p}_1^{\prime }}-E_{\mathbf{p}_2^{\prime }})$
and ``propagator'' (\ref{propagator}) is now related to the meson line in
the left graph in Fig.~4.
\begin{figure}[ht]
\scalebox{0.5}{\includegraphics{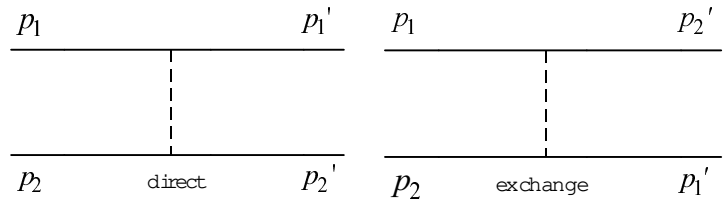}}
\caption{\label{labalfig4}
On-energy-shell contributions to the $NN$ interaction
(the $g^{2}$-order Feynman diagrams for $NN$ scattering).}
\end{figure}

One could take this analysis a little further, and show that the
off-energy-shell deviations from the conventional on-energy-shell result
obtained in the static approximation are of the order $\mathbf{p}_r^2/m^2$,
where $\mathbf{p}_r$ is the relative $NN$ momentum. In other words, the
corresponding off-energy-shell corrections have a relativistic origin.

Explicit formulae for the $NN$-quasipotential generated by the exchange of
heavier mesons are given in the next section.

\subsection{$NN \leftrightarrow \pi NN$ transition operators}

To separate the $N N \leftrightarrow \pi N N $ interaction operators one
needs to perform the normal ordering of fermion operators in the fourth and
fifth integrals of Eq. (\ref{[R,[R,V]]ExplicitForm}) and their Hermitian
conjugate parts. After a lengthy algebra we get the pion production operator,

\begin{eqnarray}
K(NN &\rightarrow &\pi NN)=\int d\mathbf{p}_1d\mathbf{p}_2d\mathbf{p}%
_1^{\prime }d\mathbf{p}_2^{\prime }d\mathbf{k}V_{\pi NN}(\mathbf{p}%
_1^{\prime },\mathbf{p}_2^{\prime },\mathbf{k};\mathbf{p}_1,\mathbf{p}_2)
\nonumber \\
&&\times a_c^{\dagger }(\mathbf{k})b_c^{\dagger }(\mathbf{p}_1^{\prime
})b_c^{\dagger }(\mathbf{p}_2^{\prime })b_c(\mathbf{p}_1)b_c(\mathbf{p}_2),
\label{PiNNInteraction}
\end{eqnarray}

\smallskip
\[
V_{\pi NN}\left( \mathbf{{p}_1^{\prime },{p}_2^{\prime },{k};{p}_1,{p}_2}%
\right) =V_{\pi NN}\left( \mathrm{Feynman-like}\right) +V_{\pi NN}\left(
\mathrm{off-energy-shell}\right) ,
\]
where 
\[
V_{\pi NN}(\mathrm{Feynman-like})=-i\frac{g^3}{(2\pi )^{9/2}}\frac{m^2}{%
\sqrt{2\omega _{\mathbf{k}}E_{\mathbf{p}_1}E_{\mathbf{p}_2}E_{\mathbf{p}%
_1^{\prime }}E_{\mathbf{p}_2^{\prime }}}}\delta (\mathbf{p}_1+\mathbf{p}_2-%
\mathbf{p}_1^{\prime }-\mathbf{p}_2^{\prime }-\mathbf{k})
\]

\begin{equation}
\times \frac{\bar{u}(\mathbf{p}_2^{\prime })\gamma _5u(\mathbf{p}_2)}{%
(p_2-p_2^{\prime })^2-\mu ^2}\bar{u}(\mathbf{p}_1^{\prime })\left[ \frac 1{%
\not{p}_1^{\prime }+\not{k}+m}+\frac 1{\not{p}_1-\not{k}+m}\right] u(\mathbf{%
p}_1),  \label{PiNNFeynman}
\end{equation}
and
\[
V_{\pi NN}(\mathrm{off-energy-shell})=-i\frac 13\frac{g^3}{(2\pi )^{9/2}}%
\frac{m^2}{\sqrt{2\omega _{\mathbf{k}}E_{\mathbf{p}_1}E_{\mathbf{p}_2}E_{%
\mathbf{p}_1^{\prime }}E_{\mathbf{p}_2^{\prime }}}}\delta (\mathbf{p}_1+%
\mathbf{p}_2-\mathbf{p}_1^{\prime }-\mathbf{p}_2^{\prime }-\mathbf{k})
\]

\[
\times \frac{\bar{u}(\mathbf{p}_2^{\prime })\gamma _5u(\mathbf{p}_2)}{%
2\omega _{\mathbf{p}_2-\mathbf{p}_2^{\prime }}}\left\{ \left( \frac 1{E_{%
\mathbf{p}_2}-E_{\mathbf{p}_2^{\prime }}+\omega _{\mathbf{p}_2-\mathbf{p}%
_2^{\prime }}}-\frac 1{E_{\mathbf{p}_1^{\prime }}+\omega _{\mathbf{k}}-E_{%
\mathbf{p}_1}+\omega _{\mathbf{p}_2-\mathbf{p}_2^{\prime }}}\right) \right.
\]
\[
\times \bar{u}(\mathbf{p}_1^{\prime })\left[ \frac 1{\not{p}_1^{\prime }+\not%
{k}+m}+\frac 1{\not{p}_1-\not{k}+m}\right] u(\mathbf{p}_1)
\]
\[
-\left( \frac 1{E_{\mathbf{p}_2}-E_{\mathbf{p}_2^{\prime }}-\omega _{\mathbf{%
p}_2-\mathbf{p}_2^{\prime }}}-\frac 1{E_{\mathbf{p}_1^{\prime }}+\omega _{%
\mathbf{k}}-E_{\mathbf{p}_1}-\omega _{\mathbf{p}_2-\mathbf{p}_2^{\prime
}}}\right)
\]

\[
\times \bar{u}(\mathbf{p}_1^{\prime })\left[ \frac 1{\not{p}_1^{\prime }+\not%
{k}+m}+\frac 1{\not{p}_1-\not{k}+m}\right] u(\mathbf{p}_1)
\]

\[
\left. +\left( \frac 1{E_{\mathbf{p}_2}-E_{\mathbf{p}_2^{\prime }}-\omega _{%
\mathbf{p}_2-\mathbf{p}_2^{\prime }}}-\frac 1{E_{\mathbf{p}_1^{\prime
}}+\omega _{\mathbf{k}}-E_{\mathbf{p}_1}-\omega _{\mathbf{p}_2-\mathbf{p}%
_2^{\prime }}}\right) \right.
\]
\[
\times \bar{u}(\mathbf{p}_1^{\prime })\left[ \frac 1{\not{p}_1+\not%
{q}+m}+\frac 1{\not{p}_1^{\prime }-\not{q}+m}\right] u(\mathbf{p}_1)
\]
\[
-\left( \frac 1{E_{\mathbf{p}_2}-E_{\mathbf{p}_2^{\prime }}+\omega _{\mathbf{%
p}_2-\mathbf{p}_2^{\prime }}}-\frac 1{E_{\mathbf{p}_1^{\prime }}+\omega _{%
\mathbf{k}}-E_{\mathbf{p}_1}+\omega _{\mathbf{p}_2-\mathbf{p}_2^{\prime
}}}\right)
\]

\begin{equation}
\times \left. \bar{u}(\mathbf{p}_1^{\prime })\left[ \frac 1{\not%
{p}_1^{\prime }+\not{q}\_+m}+\frac 1{\not{p}_1-\not{q}\_+m}\right] u(\mathbf{%
p}_1)\right\}   \label{PiNNOffshell}
\end{equation}
with $q=p_2-p_2^{\prime }$ and $q\_=(\omega _{\bf q},-{\bf q})$.
%

We have used the relation
\begin{equation}
\left[ (x - y)^2 - \mu^ 2 \right]^{-1} = \frac{1}{\omega_{\mathbf{x} -
\mathbf{y}} } \left(\left[x_0 - y_0 - \omega_{\mathbf{x} - \mathbf{y}}
\right]^{-1} - \left[x_0 - y_0 + \omega_{\mathbf{x} - \mathbf{y}%
}\right]^{-1} \right),  \label{Relation}
\end{equation}
that holds for any four-vectors $x=(x_0, \mathbf{x})$ and $y=(y_0, \mathbf{y}%
)$ with $\omega_{\mathbf{z}}=\sqrt{\mathbf{z}^2 + \mu^ 2 }$.

On the energy shell:
\begin{equation}
E_{\mathbf{p}_1}+E_{\mathbf{p}_2}=E_{\mathbf{p}_1^{\prime }}+E_{\mathbf{p}%
_2^{\prime }}+\omega _{\mathbf{k}}  \label{PiNNEnergyShell}
\end{equation}
expression (\ref{PiNNFeynman}) yields the pion production $T$ matrix in the $%
g^3$-order, i.e., to the first nonvanishing approximation in the CPR. We
remind that a clothed pion cannot be absorbed or emitted by a clothed
nucleon. At the same time the r.h.s. of (\ref{PiNNOffshell}) becomes equal
to zero since under condition (\ref{PiNNEnergyShell} ) the propagators in
its round brackets cancel each other.

By analogy with the interaction between clothed nucleons we could introduce
a quasipotential defined by the matrix elements of the operator (\ref
{PiNNInteraction}) between the properly symmetrized clothed $NN$ and $\pi NN$
states
\begin{eqnarray}
\tilde V_{\pi N N}(\mathbf{p }_1^\prime, \mathbf{p }_2^\prime, \mathbf{k } ;
\mathbf{p }_1,\mathbf{p }_2) = \left\langle a_c^\dagger ( \mathbf{k} )
b_c^\dagger ( \mathbf{p}_1^\prime ) b_c^\dagger (\mathbf{p}_2^\prime )
\Omega \mathbf{|} K(N N\rightarrow \pi N N) \mathbf{|} b_c^\dagger ( \mathbf{%
p}_1 ) b_c^\dagger ( \mathbf{p}_2 ) \Omega \right\rangle  \nonumber \\
= \frac{1}{2} \left\{ V_{\pi N N}(\mathbf{p }_1^\prime, \mathbf{p }%
_2^\prime, \mathbf{k} ; \mathbf{p }_1, \mathbf{p }_2) - V_{\pi N N}(\mathbf{%
p }_1^\prime, \mathbf{p }_2^\prime, \mathbf{k} ; \mathbf{p }_2,\mathbf{p }%
_1) \right.  \nonumber \\
\left. - V_{\pi N N}(\mathbf{p }_2^\prime, \mathbf{p }_1^\prime, \mathbf{k }%
; \mathbf{p }_1,\mathbf{p }_2) + V_{\pi N N}(\mathbf{p }_2^\prime, \mathbf{p
}_1^\prime, \mathbf{k } ; \mathbf{p }_2,\mathbf{p }_1) \right\}
\label{PiNNQuasipotential}
\end{eqnarray}
Of course, such matrix elements characterize only some part of that physical
input which contributes to pion-nuclear dynamics with the operator (\ref
{PiNNInteraction}) included in the interaction $K_I$.

All the terms involved in the \thinspace $V_{\pi NN}$ quasipotential may be
divided into the two groups. The first group refers to the pion production
mechanisms shown schematically in Fig.~5.
\begin{figure}[ht]
\scalebox{0.5}{\includegraphics{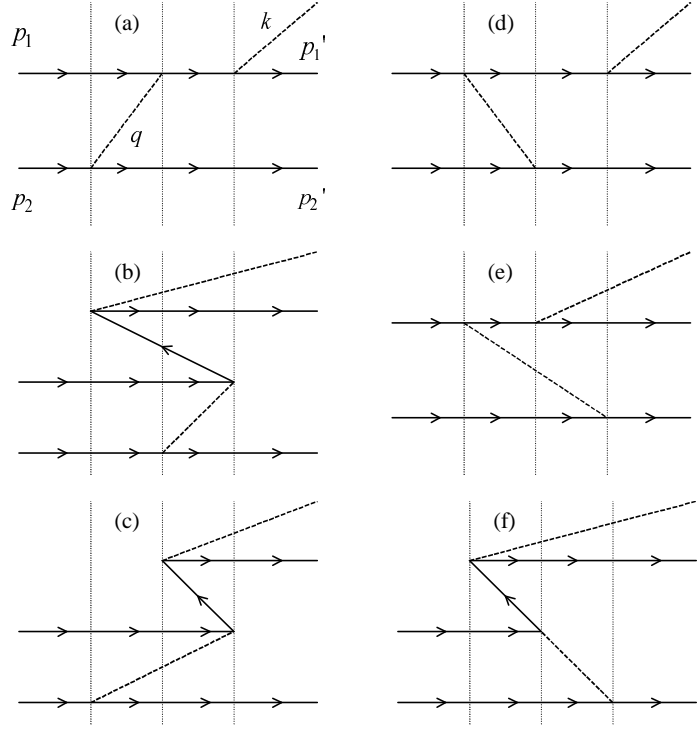}}
\caption{\label{labalfig5}
Illustration of the ``retarded'' pion production mechanisms on the
$NN$ pair in the \ $g^3$-order.}
\end{figure}

Note that the three left graphs in Fig. 5 exemplify the processes in which
the pion created by one nucleon is scattered by another one, while the three
graphs on the r.h.s. show the processes, where one nucleon emits two pions
one of which is absorbed by another nucleon. Again, on the energy shell the
sum of all these contributions can be represented by the Feynman graph
(Fig.~6, left) with the pion exchange preceding the pion emission.
\begin{figure}[ht]
\scalebox{0.5}{\includegraphics{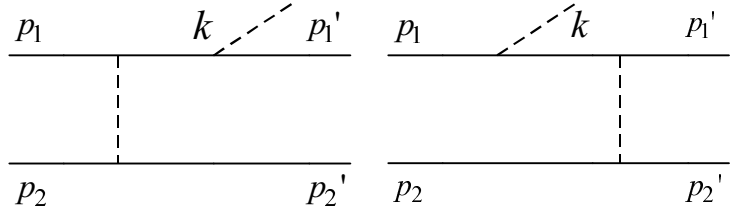}}
\caption{\label{labalfig6}
The conventional Feynman diagrams for the $NN\rightarrow \pi NN$
process in the $g^{3}$--order.}
\end{figure}

The second group of terms which could be extracted from Eqs. (\ref
{PiNNFeynman}) and (\ref{PiNNOffshell}) are shown in Fig.~7.
\begin{figure}[ht]
\scalebox{0.5}{\includegraphics{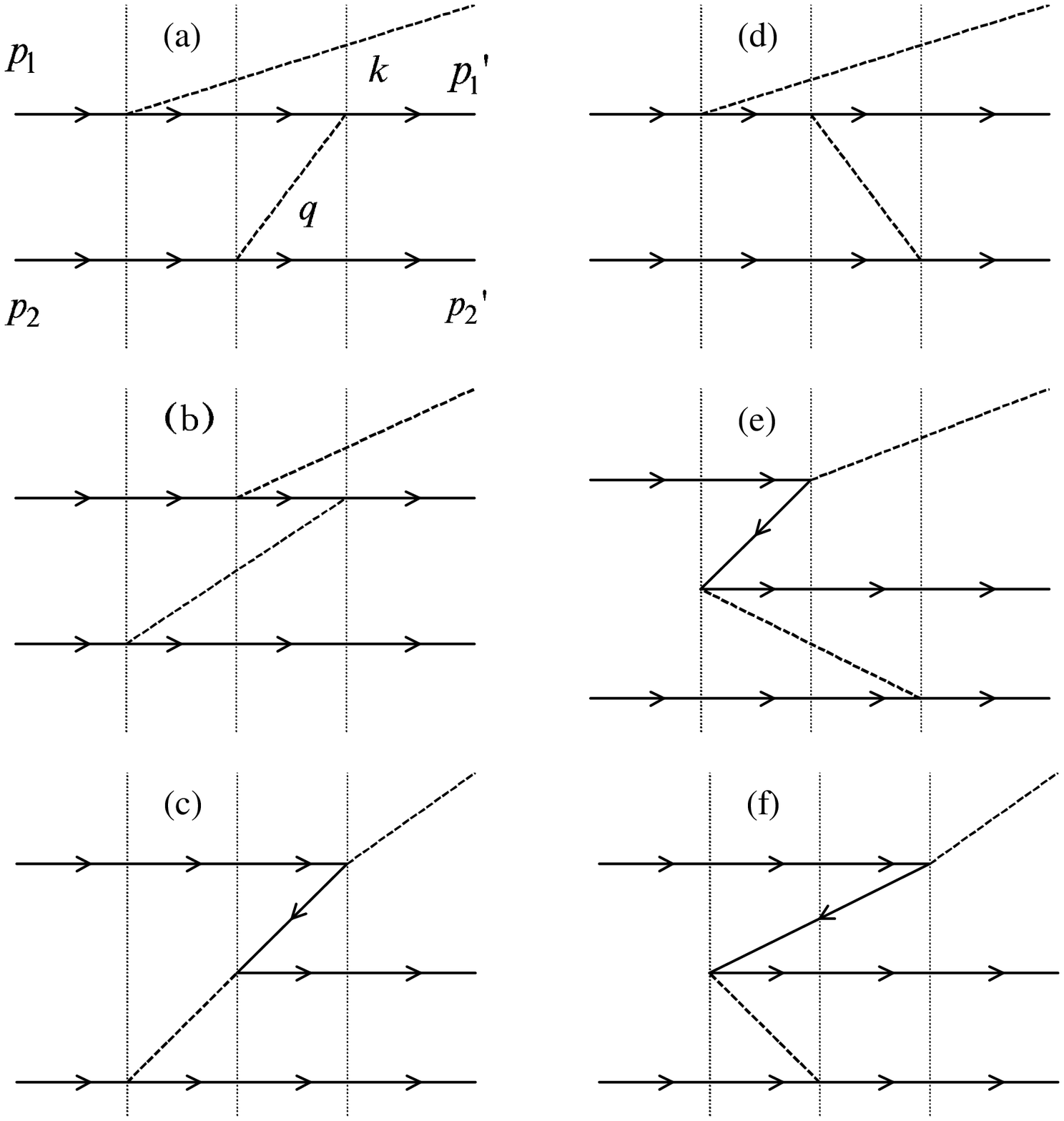}}
\caption{\label{labalfig7}
Illustration of the ``advanced'' pion production mechanisms on the
$NN$ pair in the \ $g^3$-order.}
\end{figure}

On the energy shell, the sum of these contributions can be represented by
the Feynman graph (Fig. 6, right) with the pion emission preceding the pion
exchange. Therefore, in the first nonvanishing approximation a pion can be
absorbed/emitted only by a correlated $NN$ pair through a mechanism of the $%
g^3$-order.
%

In Ref. \cite{KobayashiSatoOhtsuboProgrTheorPhys1997} the UT method has been
applied to the same $NN\rightarrow \pi NN$ process assuming a meson-nucleon
PV-coupling. Taking into account this difference in the coupling scheme and
relying upon our Eqs. (\ref{PiNNFeynman}) and (\ref{PiNNOffshell}), one can
reproduce those effective interactions given first in \cite
{KobayashiSatoOhtsuboProgrTheorPhys1997}.

In search of physical interpretation of the obtained results let us consider
the expression
\begin{equation}
\frac{\bar{u}(\mathbf{p}_2^{\prime })\gamma _5u(\mathbf{p}_2)}{E_{\mathbf{p}%
_2}-E_{\mathbf{p}_2^{\prime }}-\omega _{\mathbf{p}_2-\mathbf{p}_2^{\prime }}}%
\bar{u}(\mathbf{p}_1^{\prime })\left[ \frac 1{\not{p}_1^{\prime }+\not%
{k}+m}+\frac 1{\not{p}_1-\not{k}+m}\right] u(\mathbf{p}_1).  \label{s.1}
\end{equation}
We could write
\[
\left[ E_{\mathbf{p}_2}-E_{\mathbf{p}_2^{\prime }}-\omega _{\mathbf{p}_2-%
\mathbf{p}_2^{\prime }}\right] ^{-1}=\left[ E-E_{\mathbf{p}_1}-E_{\mathbf{p}%
_2^{\prime }}-\omega _{\mathbf{p}_2-\mathbf{p}_2^{\prime }}\right] ^{-1}
\]
with $E=E_{\mathbf{p}_1}+E_{\mathbf{p}_2}$ that is prompted with the
graphical representation of (\ref{s.1}) (more exactly, its part) in Fig.~8a,
where all vertices displayed by circles (''interaction points'') are
connected by bold lines (among them the nucleon line has the right arrow to
distinguish nucleons and antinucleons).
\begin{figure}[ht]
\scalebox{0.5}{\includegraphics{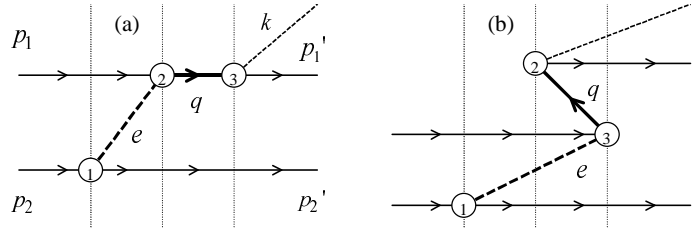}}
\caption{\label{labalfig8}
Graphical representation of Eqs. (\ref{s.1}).
}
\end{figure}

As admitted in the OFPT, the particle three-momenta are related to the lines
in Fig. 8a via the following prescription: when moving from the left to the
right, a sum of ''outgoing'' momenta is equal to a sum of ''ingoing'' ones.
Here the term ''ingoing'' (''outgoing'') is referred to the corresponding
line or lines lying on the left (right) from the dotted vertical (a
''phantom'' line) that passes through a given vertex. These vertices and
phantoms are enumerated in the alphabetical order 1, 2, 3, ... Accordingly,
we have
\begin{equation}
\mathbf{p}_2=\mathbf{e}+\mathbf{p}_2^{\prime },\,\,\,\,\,\,\,\,\,\mathbf{p}%
_1+\mathbf{e}=\mathbf{q}=\mathbf{k}+\mathbf{p}_1^{\prime }.  \label{s.2}
\end{equation}
Evidently, being taken together, these equations provide the total momentum
conservation,
\begin{equation}
\mathbf{p}_1+\mathbf{p}_2=\mathbf{k}+\mathbf{p}_1^{\prime }+\mathbf{p}%
_2^{\prime }.  \label{s.3}
\end{equation}

In parallel, we could consider Fig. 8b, where $\mathbf{p}_2=\mathbf{e}+%
\mathbf{p}_2^{\prime }$, $\mathbf{e}+\mathbf{p}_1+\mathbf{q}=0$ and $\mathbf{%
q}+\mathbf{p}_1^{\prime }+\mathbf{k}=0$. In particular, it means that $-%
\mathbf{q}=\mathbf{p}_1^{\prime }+\mathbf{k}=\mathbf{e}+\mathbf{p}_1=\mathbf{%
p}_2-\mathbf{p}_2^{\prime }$ in accordance with (\ref{s.3}).

As in case of Fig. 8a, the three internal lines between the phantoms 1 and 2
can be associated with the OFPT propagator, $\left[ E_{\mathbf{p}_2}-E_{%
\mathbf{p}_2^{\prime }}-\omega _{\mathbf{p}_2-\mathbf{p}_2^{\prime }}\right]
^{-1}$.

Regarding the lines between the phantoms 2 and 3 in Fig. 8a and Fig. 8b one
can make up the propagators (cf. discussion in Sect. 3.2)
\begin{equation}
\frac 1{E-E_{\mathbf{q}}-E_{\mathbf{p}_2^{\prime }}}|_{E=E_{\mathbf{p}%
_1^{\prime }}+E_{\mathbf{p}_2^{\prime }}+\omega _{\mathbf{k}}}=\frac 1{E_{%
\mathbf{p}_1^{\prime }}-E_{\mathbf{q}}+\omega _{\mathbf{k}}}  \label{s.1a}
\end{equation}
and
\begin{equation}
\left[ E-\omega _{\mathbf{k}}-E_{\mathbf{p}_1^{\prime }}-E_{\mathbf{q}}-E_{%
\mathbf{p}_1}-E_{\mathbf{p}_2^{\prime }}-\omega _{\mathbf{p}_2-\mathbf{p}%
_2^{\prime }}\right] ^{-1}=-\,\,\,\left[ E_{\mathbf{p}_1^{\prime }}+\omega _{%
\mathbf{k}}+E_{\mathbf{q}}\right] ^{-1}  \label{s.2a}
\end{equation}
with $E=E_{\mathbf{p}_1}+E_{\mathbf{p}_2^{\prime }}+\omega _{\mathbf{p}_2-%
\mathbf{p}_2^{\prime }}$ .

Handling the off-energy-shell graphs such as in Figs. 8a and 8b we cannot
\textit{a priori} ignore another option: with $E=E_{\mathbf{p}_1}+E_{\mathbf{%
p}_2^{\prime }}+\omega _{\mathbf{p}_2-\mathbf{p}_2^{\prime }}$ in (\ref{s.1a}%
) and $E=E_{\mathbf{p}_1^{\prime }}+E_{\mathbf{p}_2^{\prime }}+\omega _{%
\mathbf{k}}$ in (\ref{s.2a}). Both options contribute to the r.h.s. of (\ref
{PiNNOffshell}), being equal to each other on the energy shell.

Doing so and using the well-known rules, where, for instance, the vertices
with the lable 2 in Fig. 8a and Fig. 8b are equivalent to the factors
\begin{equation}
i\frac g{\sqrt{2\left( 2\pi \right) ^3\omega _{\mathbf{p}_2-\mathbf{p}%
_2^{\prime }}}}\frac m{\sqrt{E_{\mathbf{p}_1}E_{\mathbf{q}}}}\bar{u}(\mathbf{%
q})\gamma _5u(\mathbf{p}_1)  \label{s.4}
\end{equation}
and
\begin{equation}
i\frac g{\sqrt{2\left( 2\pi \right) \omega _{\mathbf{k}}}}\frac m{\sqrt{E_{%
\mathbf{p}_1^{\prime }}E_{\mathbf{q}}}}\bar{u}(\mathbf{p}_1^{\prime })\gamma
_5u(\mathbf{q}),  \label{s.5}
\end{equation}
respectively, it is easy to get the matrix element
\begin{equation}
\bar{u}(\mathbf{p}_1^{\prime })\left[ \not{p}_1^{\prime }+\not{k}+m\right]
^{-1}u(\mathbf{q})  \label{s.6}
\end{equation}
typical of the single-nucleon matrix elements involved in (\ref{PiNNFeynman}%
) and (\ref{PiNNOffshell}).

Relying upon the discussion below (\ref{PiNEnergyShell}) one may say that
the quantity (\ref{s.6}) is related to the ''right-hand'' $s$-pole mechanism
of the off-energy-shell scattering of the intermediate meson (pion), that
has been emitted by nucleon 2, on  nucleon 1.

As in Subsects. 3.2-3.3, one has to stress again that, unlike the Feynman
diagrams in Fig. 6, the meson is on its mass shell. Being a mediator in
forming different pion production mechanisms on the correlated pair of
nucleons, it is intimately embedded in the diagram of Fig. 9a,
where, first, one of the
interacting nucleons ''shakes off'' the pion (the bright spot) which is
absorbed, then, by the other nucleon (the hatched spot) with the subsequent
(''retarded'') emission of the detected meson (pion).
\begin{figure}[ht]
\scalebox{0.5}{\includegraphics{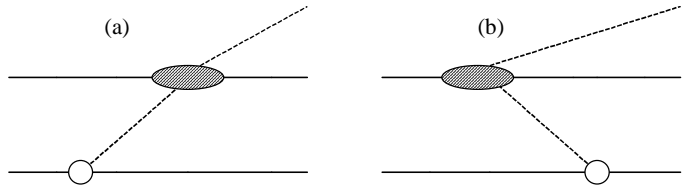}}
\caption{\label{labalfig9}
Graphical representation for the ``retarded'' (a) and ``advanced'' (b)
pion production mechanisms on the $NN$ pair.}
\end{figure}

Of course, such ''chronological'' expressions which employ the terms
''first'' and ''then'' has nothing common with any real time development of the process $%
NN\leftrightarrow \pi NN$. In fact, we imply a left-to-right-hand alignment
of intermediate pion-fermion states that occur when evaluating the matrix
elements $\left\langle \pi NN|V|n\right\rangle \left\langle
n|V|m\right\rangle \left\langle m|V|NN\right\rangle $, where $\left|
m\right\rangle $ and $\left| n\right\rangle $ are permissible $K_F$
eigenstates. In this respect, it is useful to keep in mind the
on-energy-shell relationship
\[
\left\langle \pi NN|{\frac 13}\left[ R,\left[ R,V\right] \right]
|NN\right\rangle =\left\langle \pi NN|VG_F\left( E\right) VG_F\left(
E\right) V|NN\right\rangle
\]
with the resolvent expressed by
\[
G_F\left( E\right) =\left( E-K_F\right) ^{-1}
\]
\[
E=E_{\mathbf{p}_1}+E_{\mathbf{p}_2}=E_{\mathbf{p}_1^{\prime }}+E_{\mathbf{p}%
_2^{\prime }}+\omega _{\mathbf{k}}.
\]
Nevertheless, it is convenient to refer to the reaction mechanisms
represented in Fig. 9a as ''retarded'', to differentiate them from the
''advanced'' ones displayed in Fig. 9b.

\section{Incorporation of heavy-meson exchanges}

In this section we have collected our more general evaluations of
quasipotentials for the
$NN \rightarrow NN$ and $NN \rightarrow \pi NN$ interactions that are due to
the exchange of $b = \pi, \eta, \rho, \omega, \delta$ and $\sigma$ mesons.

\subsection{Pion and heavier-meson exchanges in $NN \rightarrow NN$ process}

Here we consider the well-known Yukawa-type meson-nucleon couplings (see,
e.g., \cite{FudaZhang.Phys.Rev.C54.495.1996}) additively involved in the
primary total Hamiltonian. Application of our approach in this case results
in the following expression for the $NN\rightarrow NN$ quasipotential
\[
\tilde{V}_{NN}(\mathbf{p}_1^{\prime },\mathbf{p}_2^{\prime };\mathbf{p}_1,%
\mathbf{p}_2)=-\frac 1{2(2\pi )^3}\frac{m^2}{\sqrt{E_{\mathbf{p}_1}E_{%
\mathbf{p}_2}E_{\mathbf{p}_1^{\prime }}E_{\mathbf{p}_2^{\prime }}}}\delta (%
\mathbf{p}_1^{\prime }+\mathbf{p}_2^{\prime }-\mathbf{p}_1-\mathbf{p}_2)
\]
\begin{equation}
\times \sum_b\tilde{V}_{NN}^b(\mathbf{p}_1^{\prime },\mathbf{p}_2^{\prime };%
\mathbf{p}_1,\mathbf{p}_2).  \label{NNQuasipotentialWithHeavierMesons}
\end{equation}

The contribution to this formula from the $\delta$- and $\sigma$- mesons
exchange is given by
\[
\tilde V_{N N}^b (\mathbf{p }_1^\prime , \mathbf{p }_2^\prime ; \mathbf{p }%
_1, \mathbf{p }_2 ) = g_b^2 h_b \frac{\bar{u}(\mathbf{p }_1^{\prime } )u(%
\mathbf{p }_1 ) \bar{u}(\mathbf{p }_2^{\prime } ) u(\mathbf{p }_2 ) } {%
(p_{1}-p_{1}^{\prime})^{2} -\mu_b ^{2}}
\]
\begin{equation}
+ (1 \leftrightarrow 2, 1^\prime \leftrightarrow 2^\prime) - (1
\leftrightarrow 2) - (1^\prime \leftrightarrow 2^\prime), b = \delta,
\sigma, \,\,\, h_{\delta} = \mathbf{\tau}_1 \cdot \mathbf{\tau}_2,
h_{\sigma} = 1.  \label{NNQuasipotentialWithDelta&SigmaMesons}
\end{equation}
As usually, the Pauli vector $\mathbf{\tau }$ is the matrix in the nucleon
isospin space. The account for the $\pi$- and $\eta$- mesons exchange gives
rise to the following contribution
\[
\tilde V_{N N}^b (\mathbf{p }_1^\prime, \mathbf{p }_2^\prime ; \mathbf{p }_1,%
\mathbf{p }_2 ) = g_b^2 h_b \frac{\bar{u}(\mathbf{p }_1^{\prime } \,
r_1^\prime) \gamma_5 u(\mathbf{p }_1 \, r_1) \bar{u}(\mathbf{p }_2^{\prime }
\, r_2^\prime) \gamma_5 u(\mathbf{p }_2 \, r_2) } {(p_{1}-p_{1}^{\prime
})^{2}-\mu_b ^{2}}
\]
\begin{equation}
+ (1 \leftrightarrow 2, 1^\prime \leftrightarrow 2^\prime) - (1
\leftrightarrow 2) - (1^\prime \leftrightarrow 2^\prime), b = \pi, \eta,
\,\,\, h_{\pi} = \mathbf{\tau}_1 \cdot \mathbf{\tau}_2, h_{\eta} = 1.
\label{NNQuasipotentialWithPi&Eta}
\end{equation}

At last, the $NN$ quasipotential originating from the $\rho$- and $\omega$-
mesons exchange can be written as

\[
\tilde V_{N N}^b (\mathbf{p }_1^\prime , \mathbf{p }_2^\prime ; \mathbf{p }%
_1 ,\mathbf{p }_2 ) = g_b^2 h_b \frac{\bar{u}(\mathbf{p }_1^{\prime } )
\gamma_{\mu} u(\mathbf{p }_1 ) \Lambda_b^{\mu \nu} (p_{1}-p_{1}^{\prime })
\bar{u}(\mathbf{p }_2^{\prime }) \gamma_{\nu} u(\mathbf{p }_2 ) } {%
(p_{1}-p_{1}^{\prime})^{2}-\mu_b ^{2}}
\]
\[
+ (1 \leftrightarrow 2, 1^\prime \leftrightarrow 2^\prime) - (1
\leftrightarrow 2) - (1^\prime \leftrightarrow 2^\prime),
\]
\begin{equation}
b = \rho, \omega, \,\,\, h_{\rho} = \mathbf{\tau}_1 \cdot \mathbf{\tau}_2,
h_{\omega} = 1, \Lambda_b^{\mu \nu} (q) = -g^{\mu \nu} + (q^{\mu} q^{\nu} /
\mu_b^2).  \label{NNQuasipotentialWithRho&Omega}
\end{equation}

\subsection{Heavy-meson exchanges in $NN \leftrightarrow \pi NN$ processes}

In regards to the $NN \leftrightarrow \pi NN$ process the corresponding
quasipotential has the form
\[
\tilde V_{\pi N N} (\mathbf{p }_1^\prime \, r_1^\prime, \mathbf{p }_2^\prime
\, r_2^\prime, \mathbf{k }; \mathbf{p }_1 \, r_1,\mathbf{p }_2 \, r_2) =
\frac{im^{2}}{3(2\pi )^{9/2}}\frac{\delta (\mathbf{p }_1+\mathbf{p} _2-%
\mathbf{p}_1^{\prime }-\mathbf{p}_2^{\prime }-\mathbf{k})}{\sqrt{E_{\mathbf{p%
}_{1}}E_{\mathbf{p}_{1}^{\prime }}E_{\mathbf{p}_{2}}E_{\mathbf{p}%
_{2}^{\prime }}}}
\]
\begin{equation}
\times \sum_{b,j} \frac{1}{\sqrt{2\omega _{\mathbf{k}}^j} } \tilde V_{\pi N
N}^{b,j} (\mathbf{p }_1^\prime \, r_1^\prime, \mathbf{p }_2^\prime \,
r_2^\prime, \mathbf{k }; \mathbf{p }_1 \, r_1,\mathbf{p }_2 \, r_2).
\label{PiNNQuasipotentialWithHeavierMesons}
\end{equation}

The $\delta $- and $\sigma $- mesons exchange contribution to this formula
is given by
\[
\tilde{V}_{\pi NN}^{b,j}(\mathbf{p}_1^{\prime },\mathbf{p}_2^{\prime },%
\mathbf{k};\mathbf{p}_1,\mathbf{p}_2)=g_\pi g_b^2\bar{u}(\mathbf{p}%
_2^{\prime })u(\mathbf{p}_2)\bar{u}(\mathbf{p}_1^{\prime })
\]
\[
\times \left\{ \left[ -\frac 2{q^2-\mu _b^2}+\frac 1{(p_1^{\prime
}+k-p_1)^2-\mu _b^2}\right] \left[ \frac{t_{b,1}^j}{\not{p}_1^{\prime }+\not%
{k}+m}-\frac{t_{b,2}^j}{\not{p}_1-\not{k}-m}\right] \right.
\]
\[
\left. +\frac 1{2\omega _q^b}\left[ \frac 1{E_{\mathbf{p}_2}-E_{\mathbf{p}%
_2^{\prime }}-\omega _q^b}+\frac 1{E_{\mathbf{p}_1}-E_{\mathbf{p}_1^{\prime
}}+\omega _q^b-\omega _{\mathbf{k}}^j}\right] \left[ -\frac{t_{b,1}^j}{\not%
{p}_1+\not{q}+m}+\frac{t_{b,2}^j}{\not{p}_1^{\prime }-\not{q}-m}\right]
\right.
\]
\[
\left. +\frac 1{2\omega _q^b}\left[ \frac 1{E_{\mathbf{p}_2^{\prime }}-E_{%
\mathbf{p}_2}-\omega _q^b}+\frac 1{E_{\mathbf{p}_1^{\prime }}-E_{\mathbf{p}%
_1}+\omega _q^b+\omega _{\mathbf{k}}^j}\right] \left[ \frac{t_{b,2}^j}{\not%
{p}_1^{\prime }+\not{q}\_-m}-\frac{t_{b,1}^j}{\not{p}_1-\not{q}\_+m}\right]
\right\}
\]
\[
\times \gamma _5u(\mathbf{p}_1\,r_1)+(1\leftrightarrow 2,1^{\prime
}\leftrightarrow 2^{\prime })-(1\leftrightarrow 2)-(1^{\prime
}\leftrightarrow 2^{\prime }),
\]
\begin{equation}
b=\delta ,\sigma ,\,\,\,t_{\delta ,1}^j=\tau _1^j\mathbf{\tau }_1\mathbf{%
\tau }_2,\,\,\,t_{\delta ,2}^j=\mathbf{\tau }_2\mathbf{\tau }_1\tau
_1^j,\,\,\,t_{\sigma ,1}^j=t_{\sigma ,2}^j=1^j.
\label{PiNNQuasipotentialWithDelta&SigmaMesons}
\end{equation}

The $\pi $- and $\eta $- mesons exchange gives the contribution
\[
\tilde{V}_{\pi NN}^{b,j}(\mathbf{p}_1^{\prime }\,r_1^{\prime },\mathbf{p}%
_2^{\prime }\,r_2^{\prime },\mathbf{k};\mathbf{p}_1\,r_1,\mathbf{p}%
_2\,r_2)=g_\pi g_b^2\bar{u}(\mathbf{p}_2^{\prime }\,r_2^{\prime })\gamma _5u(%
\mathbf{p}_2\,r_2)\bar{u}(\mathbf{p}_1^{\prime }\,r_1^{\prime })
\]
\[
\times \left\{ \left[ \frac 2{q^2-\mu _b^2}+\frac 1{(p_1^{\prime
}+k-p_1)^2-\mu _b^2}\right] \left[ \frac{t_{b,1}^j}{\not{p}_1^{\prime }+\not%
{k}+m}+\frac{t_{b,2}^j}{\not{p}_1-\not{k}+m}\right] \right.
\]
\[
\left. +\frac 1{2\omega _q^b}\left[ \frac 1{E_{\mathbf{p}_2}-E_{\mathbf{p}%
_2^{\prime }}-\omega _q^b}+\frac 1{E_{\mathbf{p}_1}-E_{\mathbf{p}_1^{\prime
}}+\omega _q^b-\omega _{\mathbf{k}}^j}\right] \left[ \frac{t_{b,1}^j}{\not%
{p}_1+\not{q}+m}+\frac{t_{b,2}^j}{\not{p}_1^{\prime }-\not{q}+m}\right]
\right.
\]
\[
\left. +\frac 1{2\omega _q^b}\left[ \frac 1{E_{\mathbf{p}_2^{\prime }}-E_{%
\mathbf{p}_2}-\omega _q^b}+\frac 1{E_{\mathbf{p}_1^{\prime }}-E_{\mathbf{p}%
_1}+\omega _q^b+\omega _{\mathbf{k}}^j}\right] \left[ \frac{t_{b,2}^j}{\not%
{p}_1^{\prime }+\not{q}\_+m}+\frac{t_{b,1}^j}{\not{p}_1-\not{q}\_+m}\right]
\right\}
\]
\[
\times u(\mathbf{p}_1)+(1\leftrightarrow 2,1^{\prime }\leftrightarrow
2^{\prime })-(1\leftrightarrow 2)-(1^{\prime }\leftrightarrow 2^{\prime }),
\]
\begin{equation}
b=\pi ,\eta ,\,\,\,t_{\pi ,1}^j=\tau _1^j\mathbf{\tau }_1\mathbf{\tau }%
_2,\,\,\,t_{\pi ,2}^j=\mathbf{\tau }_2\mathbf{\tau }_1\tau
_1^j,\,\,\,t_{\eta ,1}^j=t_{\eta ,2}^j=1^j.
\label{PiNNQuasipotentialWithPi&EtaMesons}
\end{equation}

Finally, the $\pi NN$ quasipotential originating from the $\rho $- and $%
\omega $- mesons exchange is determined by
\[
\tilde{V}_{\pi NN}^{b,j}(\mathbf{p}_1^{\prime }\,r_1^{\prime },\mathbf{p}%
_2^{\prime }\,r_2^{\prime },\mathbf{k};\mathbf{p}_1\,r_1,\mathbf{p}%
_2\,r_2)=g_\pi g_b^2\bar{u}(\mathbf{p}_2^{\prime }\,r_2^{\prime })\gamma
_\mu u(\mathbf{p}_2\,r_2)\Lambda _b^{\mu \nu }(p_2-p_2^{\prime })\bar{u}(%
\mathbf{p}_1^{\prime }\,r_1^{\prime })
\]
\[
\times \left\{ \left[ \frac 2{q^2-\mu _b^2}+\frac 1{(p_1^{\prime
}+k-p_1)^2-\mu _b^2}\right] \left[ -\frac{t_{b,1}^j\gamma _\nu }{\not%
{p}_1^{\prime }+\not{k}+m}+\frac{t_{b,2}^j\gamma _\nu }{\not{p}_1-\not{k}+m}-%
\frac{2t_{b,2}^j(p_1-k)_\nu }{(p_1-k)^2-m^2}\right] \right.
\]
\[
+\frac 1{2\omega _q^b}\left[ \frac 1{E_{\mathbf{p}_2}-E_{\mathbf{p}%
_2^{\prime }}-\omega _q^b}+\frac 1{E_{\mathbf{p}_1}-E_{\mathbf{p}_1^{\prime
}}+\omega _q^b-\omega _{\mathbf{k}}^j}\right]
\]
\[
\times \left[ -\frac{t_{b,1}^j\gamma _\nu }{\not{p}_1+\not{q}+m}+\frac{%
t_{b,2}^j\gamma _\nu }{\not{p}_1^{\prime }-\not{q}+m}-\frac{%
2t_{b,2}^j(p_1^{\prime }-q)_\nu }{(p_1^{\prime }-q)^2-m^2}\right]
\]
\[
+\frac 1{2\omega _q^b}\left[ \frac 1{E_{\mathbf{p}_2^{\prime }}-E_{\mathbf{p}%
_2}-\omega _q^b}+\frac 1{E_{\mathbf{p}_1^{\prime }}-E_{\mathbf{p}_1}+\omega
_q^b+\omega _{\mathbf{k}}^j}\right]
\]
\[
\times \left. \left[ \frac{t_{b,2}^j\gamma _\nu }{\not{p}_1^{\prime }+\not%
{q}\_+m}-\frac{t_{b,1}^j\gamma _\nu }{\not{p}_1-\not{q}\_+m}-\frac{%
2t_{b,2}^j(p_1^{\prime }+q\_)_\nu }{(p_1^{\prime }+q\_)^2-m^2}\right]
\right\}
\]
\[
\times \gamma _5u(\mathbf{p}_1\,r_1)+(1\leftrightarrow 2,1^{\prime
}\leftrightarrow 2^{\prime })-(1\leftrightarrow 2)-(1^{\prime
}\leftrightarrow 2^{\prime }),
\]
\begin{equation}
b=\rho ,\omega ,\,\,\,t_{\rho ,1}^j=\tau _1^j\mathbf{\tau }_1\mathbf{\tau }%
_2,\,\,\,t_{\rho ,2}^j=\mathbf{\tau }_2\mathbf{\tau }_1\tau
_1^j,\,\,\,t_{\omega ,1}^j=t_{\omega ,2}^j=1,\Lambda _b^{\mu \nu
}(q)=-g^{\mu \nu }+(q^\mu q^\nu /\mu _b^2).
\label{PiNNQuasipotentialWithPho&OmegaMesons}
\end{equation}

\section{Summary}

We have presented a field-theoretical method to construct in a consistent way
relativistic (Hamiltonian) interactions for the meson-nucleon systems. In
particular, the interaction operators for processes of the type $%
NN\rightarrow NN$, $\pi N\rightarrow \pi N$, and $NN\leftrightarrow \pi NN$
are derived on one and the same physical footing. The method is based on the
unitary clothing approach which introduces a new representation of the
primary total Hamiltonian in terms of the operators for creation and
destruction of the so-called clothed particles, namely those particles that
can be observed. Within the approach all interactions constructed are
responsible for physical (not virtual) processes in a given system of
interacting fields. Such interactions are Hermitian and energy independent
including the off-energy-shell and recoil effects (the latter in all orders
of the $1/c^2$ - expansion). The persistent clouds of virtual particles are
no longer explicitly contained in the representation, and their effects are
included in the properties of the clothed particles and in the interactions
between them. We have also considered the additional contributions to such
interactions arising from heavy-meson exchange mechanisms. Such mechanisms
are an important subject of investigation in current research on few-body
systems, and their true role in the pion-production processes (as well as
when constructing three-nucleon forces ) has not yet been fully established.
Such issues can be best investigated, in our opinion, via the unitary
clothing transformation method which provides a well-defined, unambiguous
definition and construction of relativistic interactions between particles
with observable properties.


\end{document}